\documentclass[]{revtex4}
\usepackage[]{graphicx}
\usepackage{latexsym}
\usepackage{amssymb}

\def\lsim{\lower.5ex\hbox{$\; \buildrel < \over \sim \;$}}
\def\gsim{\lower.5ex\hbox{$\; \buildrel > \over \sim \;$}}

\def\pmb#1{\setbox0=\hbox{$#1$}%
\kern-.025em\copy0\kern-\wd0
\kern.05em\copy0\kern-\wd0
\kern-.025em\raise.0433em\box0}
\def\lsim{\lower.5ex\hbox{$\; \buildrel < \over \sim \;$}}
\def\gsim{\lower.5ex\hbox{$\; \buildrel > \over \sim \;$}}

\def\vc{\bf {{\vert_{(r=r_c)}}}}

\def\eker{$\left[{\cal E},\lambda,\gamma,a\right]$}

\begin{document}



\title{On Spin Dependence of Relativistic Acoustic Geometry}

\author{Hung-Yi Pu}
\email{hypu@phys.nthu.edu.tw}
\affiliation{Department of Physics, National Tsing Hua University, Hsinchu 30013, Taiwan}

\author{Ishita Maiti}
\email{imaitys@gmail.com}
\affiliation{Fergusson College, Pune-411 004, India}

\author{Tapas Kumar Das}
\email{tapas@mri.ernet.in}
\affiliation{S. N. Bose National Centre for Basic Sciences, Kolkata-700 098, India\\
Permanent Affiliation: Harish Chandra Research Institute, Allahabad-211 019, India.}

\author{Hsiang-Kuang Chang}
\email{hkchang@phys.nthu.edu.tw}
\affiliation{Department of Physics, National Tsing Hua University, Hsinchu 30013, Taiwan\\
Institute of Astronomy, National Tsing Hua University, Hsinchu 30013, Taiwan}

\begin{abstract}
\noindent
This work makes the first ever attempt to understand the influence of
the black hole background space-time in determining the fundamental
properties of the embedded relativistic acoustic geometry. To
accomplish such task, the role of the spin angular momentum of the
astrophysical black hole (the Kerr parameter $a$ -- a representative
feature of the background black hole metric) in estimating the value
of the acoustic surface gravity (the representative feature of the
corresponding analogue space time) has been investigated.
Since almost all astrophysical black holes
are supposed to posses some degree of intrinsic rotation,
the influence of the Kerr parameter on classical analogue
models is very important to understand.
General relativistic,
axially symmetric, non-self gravitating inflow of the hydrodynamic fluid
onto a rotating astrophysical black hole has been studied from the
dynamical systems
point of view. The location of the acoustic horizon inside such fluid flow
has been identified and the associated acoustic surface gravity has been
estimated. The dependence of such surface gravity has been studied as a
function of the Kerr parameter as well as with other dynamical and thermodynamic
variables governing the fluid flow under strong gravity. It has been
demonstrated that
for retrograde flow, the surface gravity (and hence the associated
analogue Hawking
temperature) correlates with the black hole spin in general, whereas for the
prograde flow, the surface gravity co-relates with the black hole spin for slow
to moderately rotating holes, but anti-correlates with the spin for
fast to extremely rotating holes. For certain values of the initial boundary
conditions, more than one acoustic horizons, namely two black hole type
and one white hole type, may form, where the surface gravity may become
formally infinite at the acoustic white hole. The connection between the
corresponding analogue Hawking temperature with astrophysically
relevant observables associated with the spectral signature has been discussed.
\end{abstract}
\maketitle

\section{On Analogue Gravity Phenomena and Acoustic Surface Gravity}
\label{sec1}
\noindent
Despite the remarkable resemblance in between a black hole and a 
usual thermodynamic system, black holes never radiate within the framework 
of the classical laws of physics. The introduction of the quantum mechanical 
effects radically changes the scenario -- black holes radiate due to the 
Hawking effects \cite{haw74,haw75}. 
Observational manifestation of the Hawking radiation for 
the astrophysical black holes is beyond the scope of the present day's 
experimental techniques. In addition, Hawking quanta may posses 
trans-Plankian frequencies, and physics beyond the Plank scale is yet 
to be realized within the framework of the contemporary knowledge 
of the theoretical physics. The aforementioned issues had been the prime
motivation behind looking for an analogue version of the Hawking radiation 
in the laboratory set up and the formulation of the analogue gravity 
phenomena was thus introduced by establishing the profound similarities 
in between the propagating perturbation within an inhomogeneous dynamical 
continuum and certain kinematical features of the space time as 
conceived in the general theory of relativity \cite{unr81,unr95,vis98,nov02,car05,bar05}. 

Contemporary research in this direction has gained widespread popularity since 
it provides a possibility of verifying certain exotic features of the 
black hole physics (as manifested in the close proximity of the event horizon,
including the Hawking radiation) directly some experimentally conceivable 
physical systems in the laboratory set up. Also since the smallest length scale 
encountered in the analogue gravity models are trans-Bohrian rather than
trans-Plankian, quantum gravity effects usually significant beyond the 
plank scale may also show up on considerably larger length within the 
framework of the analogue gravity formalism. 

For the common-most example where the aforementioned perturbation is 
acoustic in nature, and the spacetime describing the fluid flow embedding 
such perturbation is Minkowskian, the acoustic surface gravity $\kappa$
can be computed as \cite{vis98}:
\begin{equation}
\kappa=\left[c_s\frac{\partial}{\partial{\eta}}\left(c_s-u\right)\right]_
{\rm Evaluated~ at~ the~ acoustic~ horizon}
\label{eq1}
\end{equation}
where $c_s$ is the position dependent sound speed (the velocity of the 
propagation of the embedded perturbation in general), $u$ is the 
bulk velocity of the fluid normal to the acoustic horizon and 
$\frac{\partial}{\partial{\eta}}$ represents the derivative taken 
along the direction normal to the acoustic horizon. 

In this context it is to 
be pointed out that for the flat Minkowskian space time describing the 
fluid flow under the influence of the Newtonian gravity as well as for 
its relativistic generalization \cite{bil99,abr06}, 
the acoustic perturbation propagates along the time like curves. 
Phonons construct the geodesics of acoustic propagation which are null
with respect to the acoustic metric (the metric governing the propagation 
of the perturbation), and generate a null surface -- the acoustic horizon.
Such a horizon forms at the transonic point of the fluid flow. Hence the 
sonic surface is identical with the acoustic horizon, and the supersonic 
flow resembles the acoustic ergo region. It is evident from eq. (\ref{eq1}) 
that the location of the acoustic/analogue\footnote{Hereafter the phrases
`acoustic' and `analogue' will be used synonymously for the sake of 
brevity.} horizon and the values of the speed of the propagation of the 
perturbation, the bulk velocity of the embedding continuum,
and their space gradients -- all evaluated on such a horizon -- will 
provide the complete knowledge of the estimate of the analogue surface gravity.
A more comprehensive expression for the acoustic surface gravity for 
a generalized analogue system (including the relativistic generalization)
may be framed as \cite{bil99,abr06,bil07}:
\begin{equation}
\kappa=\left|\frac{\sqrt{{\chi^\mu}{\chi_\mu}}}{\left(1-{c_s}^2\right)}
\frac{\partial}{\partial{\eta}}\left(u-{c_s}\right)\right|_
{\rm Evaluated~ at~ the~ acoustic~ horizon}
\label{eq2}
\end{equation}
where $\chi^\mu$ is the Killing field which is null on the corresponding 
acoustic horizon and 
$\frac{\partial}{\partial{\eta}}=\eta^\mu{\partial_\mu}$. 
The algebraic expression corresponding to the $\sqrt{\chi^\mu{\chi_\mu}}$
can be evaluated once the space time metric describing the fluid flow 
as well as the propagation of the perturbation in a specified geometry 
is defined. In this context, eq. (\ref{eq2})
will further be discussed in greater detail in the subsequent sections.

\section{Accreting Astrophysical Black Holes as Analogue Systems}
\label{sec2}
Conventional works on the classical analogue gravity phenomena focus mainly 
on physical systems not directly subjected to the gravitational force.
Gravity like effects rather appear to be an emergent phenomena in such 
configurations. In such systems, only the analogue Hawking effects may be 
observed and no source for the conventional Hawking radiation may 
therefore be realized. Quite recently, attempts have been made to study 
the analogue gravity phenomena in astrophysical systems where the strong 
gravity plays a significant role to influence the relevant features 
of the acoustic geometry for Newtonian, semi Newtonian and complete 
general relativistic flows, respectively
\cite{das04,kin04,das05,abr06,nas07,das07,mac08,mac09,nag12}
\footnote{It is interesting to note that
from the historical viewpoint,
however, the first ever study on the analogue system was perhaps 
accomplished by Moncrief (1980)\cite{mon80} which was essentially performed 
on the accreting astrophysical flows.}. 
Special emphasis has been put to study such effects for accreting 
astrophysical black holes. Transonic accretion onto galactic and 
extra galactic black holes has recently been shown to be a very interesting 
example of the classical analogue systems found naturally in the 
Universe \cite{das04,das05,abr06,das07}. Such 
systems are unique in the sense that only for such configuration, both the
gravitational as well as the acoustic horizons exist simultaneously.
This makes the accreting black hole candidates the only analogue 
systems found in the Universe so far where the same flow encounters both
kind of horizons. Such systems allow to perform 
a comprehensive study of the properties of space time in the close 
proximity of both the horizons. 

In this context, spherically symmetric as well as the axisymmetric 
accretion configuration has been studied. The non linear equations describing 
the stationary, inviscid, hydrodynamic accretion onto astrophysical 
black holes may be tailored to form a first order autonomous dynamical 
systems \cite{ray02,afs03,ray03a,ray03b,ray05,ray07,bha09}. 
The physical transonic solutions
for such configuration may formally be realized as critical solutions on the 
radial Mach number versus radial distance phase portrait of the flow.
For low angular momentum 
sub-Keplerian axisymmetric accretion of inviscid hydrodynamic fluid, 
multiple critical points 
(at most three) may appear and the the integral flow solution may contain 
stationary shock \cite{lia80,abr81,muc82,muc83,muc86,fuk83,lu86,lu86,fuk87,cha89,kaf94,yan95,lu97,das12}. Any real physical global transonic solution 
joining the event horizon with the source of the accreting material will 
perforce have to pass through a saddle type critical point. 
Conventionally, a sonic point is defined at the radial distance where the 
radial Mach number becomes unity, i.e., $u$ becomes exactly equal to $c_s$. 

Depending on the equation of state used to describe the flow and 
the flow geometry, a critical point may or may not coincide with a 
sonic point. For axisymmetric accretion flow governed by the adiabatic 
equation of state, critical points are not isomorphic with the sonic points
when the flow configuration is assumed to be in vertical equilibrium
\cite{das07,das12}. A constant 
height flow or flow in conical equilibrium produces critical points which 
are identical with the sonic points when the adiabatic equation of state is 
used \cite{abr06,nag12}. 
For isothermal flow, critical points and sonic points are found to be 
identical irrespective of the flow geometry used 
\cite{nag12}. Hence for adiabatic axisymmetric accretion in vertical 
equilibrium, the critical surface can not be considered as the acoustic 
horizon, rather the location of the acoustic horizon has to be computed 
by numerically integrating the differential equations 
describing the accretion solution starting from the critical point -- 
detail procedure for which will further be discussed in greater detail 
in the subsequent sections.  

Usually for the low angular momentum accretion studied in this work, a complete 
multi-transonic solution may be referred to the configuration where 
two transonic solutions through two different saddle type critical/sonic
points are connected through a discontinuous shock transition. Such a 
stationary shock usually formed in between the outer saddle type sonic point 
and the middle centre type sonic point. A comprehensive discussion 
about such flow configuration is available in \cite{das07} \& \cite{das12}.

From the analogue gravity point of view, the aforementioned multi-transonicity
may be realized through the presence of more than one acoustic horizon of 
different kind. As obvious that the acoustic horizon may be defined by the 
equation $u^2-c_s^2=0$, for stationary axisymmetric flow configuration 
the acoustic horizon on the equatorial plane is a stationary circle 
of fixed radius. The measure of such radius depends on the initial boundary 
condition describing the accretion flow. Once the corresponding acoustic metric
$G_{\mu\nu}$ is appropriately defined over the stationary background metric 
$g_{\mu\nu}$, the discriminant of $G_{\mu\nu}$  as defined by 
${\cal D}=G^2_{t\phi}-G_{tt}G_{\phi\phi}$ may be used as the marker for
categorizing various sonic states of the flow, and can as well be used 
to determine whether the acoustic horizon is of black hole or of 
white hole type. Supersonic (subsonic) flow is characterized by the 
positivity (negativity) of ${\cal D}$ \cite{abr06,das07},
 and the change of sign of ${\cal D}$ occurs
at the acoustic horizon. Hence the sonic points are characterizes by 
${\cal D} <0 \longrightarrow {\cal D} > 0$ transition whereas
${\cal D} >0 \longrightarrow {\cal D} < 0$ marks the presence of a stationary 
shock. Following the classification of Barcel\'o, Liberati, Sonego 
\& Visser (2004), it can be shown that for certain values of the 
initial boundary conditions describing the accretion flow, two 
acoustic black holes may form at the two sonic points and an acoustic 
white hole may form at the shock location flanked in between two such 
sonic points. The acoustic surface gravity becomes formally infinite at such 
white holes because both $c_s$ and $u$ changes discontinuously at the 
shock location.
\section{Connection between the acoustic surface gravity and black hole 
spin}
\label{sec3}
\noindent
In this work, we intend to study the acoustic geometry of an axially 
symmetric accretion flow in the Kerr metric for a configuration 
more complex and astrophycially relevant compared to the idealized disc model 
of uniform thickness as studied in \cite{abr06}. 
Our present work explores the low angular momentum relativistic accretion 
in vertical equilibrium (i.e., where the flow thickness is an analytically 
defined function of radial distance, as introduced in \cite{das12} 
using the dynamical systems approach).The essential motivation will be to 
investigate how the spin angular momentum of the astrophysical 
black holes influences the estimation of the acoustic surface gravity 
for multiple acoustic horizons. 
This is to understand how the properties of the curved space time 
governing the fluid flow (the Kerr parameter $a$) determines the 
basic features of the relevant acoustic metric (the corresponding acoustic 
surface gravity $\kappa$). Such information will allow to probe a 
physical phenomena of profound significance -- the role of the background 
black hole metric (as experienced by the embedding fluid flow) in constructing
the perturbative curved manifold determining the properties of the space time 
at the close proximity of the acoustic horizon. Hence the nature of the 
dependence of the value of the acoustic surface gravity at the 
transonic surface of the accreting fluid on the Kerr parameter will 
manifest the interwining between the black hole metric and the 
acoustic metric. 

Along with the study of the black hole spin dependence of $\kappa$, we would 
also like to investigate whether a physical quantity as abstract as the 
analogue surface gravity may be associated with some observable 
phenomena in connection to the black hole astrophysics. As has already been 
explained, the low angular momentum accretion flow in the Kerr metric may generate 
two acoustic black holes at its two sonic points and an acoustic 
white hole at the shock location. At the shock, the accreting fluid 
will be heated up and will get considerably compressed. The amount of
compression will depend on the strength of the shock, where the shock 
strength $R_M=\frac{M_-}{M_+}$ is conventionally defined as the ratio of the 
pre- to the post-shock radial Mach number. Eventually, $\frac{M_-}{M_+}$
depends on the initial boundary conditions, including the black hole spin.
The shock strength $R_M$ is a significant marker of the spectral
features of the accreting black hole systems 
\cite{cha95,shr98,lau99,bor99}. Stronger is the shock, 
prominent is the spectral part indicating the radiation from the post shock 
flow. Hence $R_M$ is a good marker for the observational 
manifestation in the shock formation for sub-Keplarian black hole accretion.
In this work we intend to study the co-relation between the value of the 
acoustic surface gravity with the shock strength (which will be different for 
different values of the initial boundary conditions governing the flow, 
including the Kerr parameter) to associate the concept of acoustic 
surface gravity with the spectral signature of the rotating black holes. 

The plan of the paper is as follows: In next section we will provide a 
brief account of the transonic black hole accretion system from the 
dynamical systems point of view and the details of such accretion in the Kerr
metric will be provided. Section \ref{sec5} will be devoted to the 
calculation of the acoustic surface gravity $\kappa$. In section \ref{sec6} 
the dependence of $\kappa$ on $a$ will clarified in detail, and the 
corresponding connection with the spectral signature will be described
by providing the dependence of $\kappa$ on various shock related 
quantities. Finally in section \ref{sec7} we will conclude.  
\section{Background Fluid Flow Configuration}
\label{sec4}
\noindent
The acoustic geometry corresponding to the stationary solution of
transonic, non self-gravitating, axisymmetric, inviscid hydrodynamic
accretion of weakly rotating sub-Keplarian compressible inhomogeneous
fluid onto a spinning black hole is considered on the background geometry in 
the Boyer-Lindquist \cite{boy67} co-ordinate. The stationarity 
and the axial symmetry correspond to the two generators of the temporal 
and the axial isometries, respectively -- leading to the fact that 
the total specific energy and angular momentum remains conserved along the 
streamline. The particular flow model for which the stationary 
axisymmetric solution of the energy momentum and the baryon number 
conservation equations is sought on the equatorial plane, is described 
in \cite{das12} in great detail. In this section such flow 
configuration will first be be introduced in brief for completeness, then 
the transonic structure for such configuration will be discussed at length. 
For further clarification regarding the transonic properties of 
general relativistic axially symmetric black hole 
accretion flow in general, the author may refer to, e.g., 
\cite{nov73,gam98,bar04,cze12}. 

\subsection{Metric Elements and Related Quantities}
\label{sec4.1}
\noindent
In general any radial distance on the equatorial plane has been scaled in the
units of $GM_{BH}/c^2$ and any velocity involved has been scaled by the 
velocity of light in vacuum, $c$. $M_{BH}$ is the mass of the black hole 
considered, and $G=c=M_{BH}=1$ is used. $-+++$ signature is used along with 
a azimuthally Lorentz boosted orthonormal tetrad basis co-rotating 
with the accreting fluid. The energy momentum tensor 
$T^{\mu\nu}=\left({\epsilon}+p\right)v^\mu{v^\nu}+pg^{\mu\nu}$ of a perfect 
fluid may be defined where $\epsilon,p$ and $v^\mu$ are the total 
mass energy density, isotropic pressure in the rest frame, and the 
four velocity of the accreting fluid, respectively. If $\rho$ is the 
corresponding rest mass density, the stationary solution of the energy 
momentum conservation equation $T^{\mu\nu}_{;\mu}=0$ and the 
baryon number conservation equation $\left({\rho}v^\mu\right)_{;\mu}=0$ 
will provide two first integrals of motion, e.g., the conserved specific 
energy ${\cal E}$ (the relativistic Bernoulli's constant) and the mass
accretion rate ${\dot M}$, respectively. The actual expression for 
${\cal E}$ and ${\dot M}$ will depend on the specific form of the metric
(and related co ordinate used), the geometric flow configuration, and the
equation of state used to describe the accretion flow. 

As for the space time metric, Boyer-Lindquist line element on the equatorial 
plane can be expressed as \cite{nov73}
\begin{equation}
ds^2=g_{{\mu}{\nu}}dx^{\mu}dx^{\nu}=-\frac{r^2{\Delta}}{A}dt^2
+\frac{A}{r^2}\left(d\phi-\omega{dt}\right)^2
+\frac{r^2}{\Delta}dr^2+dz^2
\label{eq3}
\end{equation}
where 
\begin{equation}
\Delta=r^2-2r+a^2, A=r^4+r^2a^2+2ra^2,\omega=2ar/A
\label{eq4}
\end{equation}
$a$ being the Kerr parameter related to the black holes spin angular momentum.
The required metric elements are:
\begin{equation}
g_{rr}=\frac{r^2}{A},~g_{tt}=
\left(\frac{A\omega^2}{r^2}-\frac{r^2{\Delta}}{A}\right),~
g_{\phi\phi}=\frac{A}{r^2},~ g_{t\phi}=g_{\phi{t}}=-\frac{A\omega}{r^2}
\label{eq5}
\end{equation}
The normal derivative ${\partial}/{\partial}{\eta}{\equiv}
{\eta^\mu}{\partial_{\mu}}=\frac{1}{\sqrt{g_{rr}}}d/dr$. The specific angular 
momentum $\lambda$ (angular momentum per unit mass) 
and the angular velocity $\Omega$ can thus be expressed as
\begin{equation}
\lambda=-\frac{v_\phi}{v_t}; \;\;\;\;\;
\Omega=\frac{v^\phi}{v^t}
=-\frac{g_{t\phi}+\lambda{g}_{tt}}{{g_{\phi{\phi}}+\lambda{g}_{t{\phi}}}}\, ,
\label{eq6}
\end{equation}
We also define 
\begin{equation} 
B=g_{\phi\phi}+2\lambda{g_{t\phi}}+\lambda^2{g_{tt}}
\label{eq7}
\end{equation}
which will be used in the subsequent section to calculate the 
value of the acoustic surface gravity.  

The flow is assumed to possess finite radial three velocity $u$ 
(will be designated as the `advective velocity') on the equatorial 
plane. Considering $v$ to be the magnitude of the three velocity, $u$ 
is the component of three velocity perpendicular to the set of timelike 
hypersurfaces $\left\{\Sigma_v\right\}$ defined by $v^2={\rm constant}$. 
The normalization criteria $v^\mu{v}_\mu=-1$ leads to the 
expression for the temporal component $v_t$ of the four velocity
\begin{equation}
v_t=
\left[\frac{Ar^2\Delta}
{\left(1-u^2\right)\left\{A^2-4\lambda arA
+\lambda^2r^2\left(4a^2-r^2\Delta\right)\right\}}\right]^{1/2} .
\label{eq8}
\end{equation}
 
Whereas the advective dynamics is assumed to be confined on the 
equatorial plane only, the thermodynamic profile of the flow is essentially 
obtained by vertically averaging the thermodynamic quantities over the 
radius dependent vertical thickness 
\begin{equation}
H(r)=\sqrt{\frac{2}{\gamma + 1}} r^{2} \left[ \frac{(\gamma - 1)c^{2}_{s}}
{\{\gamma - (1+c^{2}_{s})\} \{ \lambda^{2}v_t^2-a^{2}(v_{t}-1) \}}\right] ^{\frac{1}{2}} ,
\label{eq9}
\end{equation}of the flow. 

The adiabatic equation of state of the form 
\begin{equation}
p=K\rho^\gamma
\label{eq10}
\end{equation}
has been used to describe the flow, where $\gamma=c_p/c_v$, the ratio 
of the specific heat at the constant pressure and at constant volume,
respectively.
\subsection{Conservation Equations and the Critical Point Conditions}
\label{sec4.2}
\noindent
The two first integrals of motion, the conserved specific energy 
${\cal E}$ of the flow and the mass 
accretion rate ${\dot M}$ can be computed as  
\begin{equation}
{\cal E} =
\left[ \frac{(\gamma -1)}{\gamma -(1+c^{2}_{s})} \right]
\sqrt{\left(\frac{1}{1-u^{2}}\right)
\left[ \frac{Ar^{2}\Delta}{A^{2}-4\lambda arA +
\lambda^{2}r^{2}(4a^{2}-r^{2}\Delta)} \right] } 
\label{eq11}
\end{equation}
\begin{equation}
{\dot M}=4{\pi}{\Delta}^{\frac{1}{2}}H(r){\rho}\frac{u}{\sqrt{1-u^2}} \, ,
\label{eq12}
\end{equation}
by integrating the stationary part of the energy momentum conservation equation 
and the continuity equation, respectively.
The corresponding entropy accretion rate ${\dot \Xi}$ that is 
conserved for the shock free polytropic accretion and increases 
discontinuously at the shock (if forms), can be expressed as
\begin{equation}
{\dot \Xi}
 = \left( \frac{1}{\gamma} \right)^{\left( \frac{1}{\gamma-1} \right)}
4\pi \Delta^{\frac{1}{2}} c_{s}^{\left( \frac{2}{\gamma - 1}\right) } \frac{u}{\sqrt{1-u^2}}\left[\frac{(\gamma -1)}{\gamma -(1+c^{2}_{s})}
\right] ^{\left( \frac{1}{\gamma -1} \right) } H(r)
\label{eq13}
\end{equation}
The conservation equations for ${\cal E}, {\dot M}$ and
${\dot \Xi}$ may simultaneously be solved 
to obtain the complete accretion profile from the dynamical systems 
point of view (see, e.g., \cite{gos07} for further detail).

The relationship between the space gradient of the acoustic velocity and 
that of the dynamical velocity can now be computed by differentiating 
eq. (\ref{eq13})
\begin{equation}
\frac{dc_s}{dr}=
\frac{c_s\left(\gamma-1-c_s^2\right)}{1+\gamma}
\left[
\frac{\chi{\psi_a}}{4} -\frac{2}{r}
-\frac{1}{2u}\left(\frac{2+u{\psi_a}}{1-u^2}\right)\frac{du}{dr} \right]
\label{eq14}
\end{equation}
whereas $du/dr$ may be obtained by differentiating eq. (\ref{eq11}) along with 
taking help of eq. (\ref{eq14})
\begin{equation}
\frac{du}{dr}=
\frac{\displaystyle
\frac{2c_{s}^2}{\left(\gamma+1\right)}
  \left[ \frac{r-1}{\Delta} + \frac{2}{r} -
         \frac{v_{t}\sigma \chi}{4\psi}
  \right] -
  \frac{\chi}{2}}
{ \displaystyle{\frac{u}{\left(1-u^2\right)} -
  \frac{2c_{s}^2}{ \left(\gamma+1\right) \left(1-u^2\right) u }
   \left[ 1-\frac{u^2v_{t}\sigma}{2\psi} \right] }}
\label{eq15}
\end{equation}
 where 
\begin{eqnarray}
\psi=\lambda^2{v_t^2}-a^2\left(v_t-1\right)~
\psi_a=\left(1-\frac{a^2}{\psi}\right)~
\sigma = 2\lambda^2v_{t}-a^2
~
& & \nonumber \\
\chi =
\frac{1}{\Delta} \frac{d\Delta}{dr} +
\frac{\lambda}{\left(1-\Omega \lambda\right)} \frac{d\Omega}{dr} -
\frac{\displaystyle{\left( \frac{dg_{\phi \phi}}{dr} + \lambda \frac{dg_{t\phi}}{dr} \right)}}
     {\left( g_{\phi \phi} + \lambda g_{t\phi} \right)}
\label{eq16}
\end{eqnarray}
Eq. (\ref{eq15}) as well as eq. (\ref{eq14}) can readily be identified with a 
set of non-linear first order differential equations representing 
autonomous dynamical systems, and their integral solutions will provide
phase trajectories on the radial Mach number M (where $M=u/c_s$) vs $r$ 
plane. The critical point condition for these integral solutions may be 
obtained by simultaneously making the numerator and the denominator of 
eq. (\ref{eq15}) vanish, and the aforementioned critical point 
condition may thus be expressed as
\begin{equation}
{c_{s}}_{\vc}={\left[\frac{u^2\left(\gamma+1\right)\psi}
                                  {2\psi-u^2v_t\sigma}
                       \right]^{1/2}_{\vc}  },
~~u{\vc}= {\left[\frac{\chi\Delta r} {2r\left(r-1\right)+ 4\Delta} \right]
^{1/2}_{\rm r=r_c}  }
\label{eq17}
\end{equation}

It is to be noted that eq. (\ref{eq17}) provides the critical point 
condition but not the location of the critical point(s). It is 
necessary to solve eq. (\ref{eq11}) under the critical point condition for 
a set of initial boundary conditions as defined by 
$\left[{\cal E},\lambda,\gamma,a\right]$. The value of $c_s$ and $u$, as 
obtained from eq. (\ref{eq17}), may be substituted at eq. (\ref{eq11}) to 
obtain a complicated non-polynomial algebraic expression for 
$r=r_c$, $r_c$ being the location of the critical point. A particular 
set of values of $\left[{\cal E},\lambda,\gamma,a\right]$ will then 
provide the numerical solution for expression to obtain 
the exact value of $r_c$. It is thus important to know the astrophycially 
relevant domain of numerical values corresponding to ${\cal E},\lambda,\gamma$ 
and $a$. 
\subsection{Astrophysically Relevant Domain of the Initial Boundary Conditions}
\label{sec4.3}
${\cal E}$ is scaled by the rest mass energy and includes the rest mass 
energy itself, hence ${\cal E}=1$ corresponds to a flow with zero 
thermal energy at infinity, which is obviously not a realistic initial 
boundary condition to generate the acoustic perturbation. Similarly,
${\cal E} <1$ is also not quite a good choice since such configuration 
with the negative energy accretion state requires a mechanism for 
dissipative extraction of energy to obtain a positive energy 
solution\footnote{A positive Bernoulii's constant flow is essential 
to study the accretion phenomena so that it can incorporate the 
accretion driven outflows ( \cite{das99} and references therein).}.
Any such dissipative mechanism is not preferred to study the acoustic 
geometry since dissipative terms in the energy momentum conservation 
equation may violate the Lorentzian invariance 
On the other hand, almost all ${\cal E}>1$ 
solutions are theoretically allowed. However, large values of 
${\cal E}$ represents accretion with unrealistically hot flows in 
astrophysics. In particular, ${\cal E}>2$ corresponds to with extremely 
large initial thermal energy which is not quite commonly observed in 
accreting black hole candidates. We thus set $1{\lsim}{\cal E}{\lsim}2$. 

A somewhat intuitively obvious range for $\lambda$ for our 
purpose is $0<\lambda{\le}2$,
since $\lambda=0$ indicates spherically symmetric flow and for $\lambda>2$
multi-critical behaviour does not show up in general. 

$\gamma=1$ corresponds to isothermal accretion where the acoustic perturbation 
propagates with position independent speed. $\gamma<1$ is not a 
realistic choice in accretion astrophysics. $\gamma>2$ corresponds to
the superdense matter with considerably large magnetic field and a 
direction dependent anisotropic pressure. The presence of a dynamically 
important magnetic field requires the solution of general relativistic 
magneto hydrodynamics equations which is beyond the scope of the present work.
Hence a choice for $1{\lsim}\gamma{\lsim}2$ seems to 
be appropriate. However, preferred bound for realistic black hole 
accretion is from $\gamma=4/3$ (ultra-relativistic flow) to $\gamma=5/3$ 
(purely non relativistic flow), see, e.g., \cite{fra02} 
for further detail. Thus we mainly concentrate on $4/3{\le}\gamma{\le}5/3$. 

The domain for $a$ lies clearly in between the values of the Kerr 
parameters corresponding to the maximally rotating black hole for 
the prograde and the retrograde flow. Hence the obvious choice 
for $a$ is $-1{\le}a{\le}1$. 
The allowed domains for the four parameter initial boundary conditions are 
thus 
$\left[1{\lsim}{\cal E}{\lsim}2, 0<\lambda{\le}2,4/3{\le}\gamma{\le}5/3,-1{\le}a{\le}1\right]$.
The aforementioned four parameters may further be classified
intro three different categories, according to the way they influence the
characteristic properties of the accretion flow. 
$\left[{\cal E},\lambda,\gamma\right]$ characterizes
the flow, and not the space–time since the accretion is assumed to
be non-self-gravitating.
The Kerr parameter $a$ exclusively
determines the nature of the space–time and hence can be thought of
as some sort of `inner boundary condition’ in qualitative sense\footnote{The effect of 
gravity is determined within the full general relativistic
framework only up to several gravitational radii. Beyond a certain
length-scale it asymptotically follows the Newtonian regime.}. 
$\left[{\cal E}, \lambda\right]{\subset}\left[{\cal E}, \lambda,\gamma\right]$ 
determines the dynamical aspects of the
flow, whereas $\gamma$ determines the thermodynamic properties. 
To follow a holistic approach, one needs to study the variation of the
salient features of the acoustic geometry on all of these four parameters.
\subsection{The Critical Velocity Gradients}
\label{sec4.4}
\noindent
Once the value of $r_c$ is computed for an astrophysically relevant 
set of \eker, the nature of the critical point(s) can also be studied 
to confirm whether it is a saddle type or a centre type critical 
point (see, e.g., \cite{gos07} for an analytical 
scheme developed using the eigenvalue problem
to accomplish such classification for axisymmetric 
accretion in the Kerr metric). The space gradient for the 
advective flow velocity at the critical point 
can also be
computed by solving the following quadratic equation 
\begin{equation}
\alpha \left(\frac{du}{dr}\right)_{\vc}^2 + 
\beta \left(\frac{du}{dr}\right)_{\vc} + \zeta = 0,
\label{eq18}
\end{equation}
where the respective co-efficients, all evaluated at the critical point $r_c$,
are obtained as
\begin{eqnarray}
\alpha=\frac{\left(1+u^2\right)}{\left(1-u^2\right)^2} - \frac{2\delta_1\delta_5}{\gamma+1}, 
 \quad \quad \beta=\frac{2\delta_1\delta_6}{\gamma+1} + \tau_6,
 \quad \quad \zeta=-\tau_5;
& & \nonumber \\
\delta_1=\frac{c_s^2\left(1-\delta_2\right)}{u\left(1-u^2\right)}, \quad \quad
\delta_2 = \frac{u^2 v_t \sigma}{2\psi}, \quad \quad
\delta_3 = \frac{1}{v_t} + \frac{2\lambda^2}{\sigma} - \frac{\sigma}{\psi} ,
\quad \quad \delta_4 = \delta_2\left[\frac{2}{u}+\frac{u v_t \delta_3}{1-u^2}\right],
& & \nonumber \\
~
\delta_5 = \frac{3u^2-1}{u\left(1-u^2\right)} - \frac{\delta_4}{1-\delta_2} -
           \frac{u\left(\gamma-1-c_s^2\right)}{a_s^2\left(1-u^2\right)},
\quad \quad \delta_6 = \frac{\left(\gamma-1-c_s^2\right)\chi}{2c_s^2} +
           \frac{\delta_2\delta_3 \chi v_t}{2\left(1-\delta_2\right)},
& & \nonumber \\
\tau_1=\frac{r-1}{\Delta} + \frac{2}{r} - \frac{\sigma v_t\chi} {4\psi},
\quad \quad
\tau_2=\frac{\left(4\lambda^2v_t-a^2\right)\psi - v_t\sigma^2} {\sigma \psi},
& & \nonumber \\
\tau_3=\frac{\sigma \tau_2 \chi} {4\psi},
\quad \quad
\tau_4 = \frac{1}{\Delta} 
       - \frac{2\left(r-1\right)^2}{\Delta^2}
       -\frac{2}{r^2} - \frac{v_t\sigma}{4\psi}\frac{d\chi}{dr},
& & \nonumber \\
\tau_5=\frac{2}{\gamma+1}\left[c_s^2\tau_4 -
     \left\{\left(\gamma-1-c_s^2\right)\tau_1+v_tc_s^2\tau_3\right\}\frac{\chi}{2}\right]
   - \frac{1}{2}\frac{d\chi}{dr},
& & \nonumber \\
\tau_6=\frac{2 v_t u}{\left(\gamma+1\right)\left(1-u^2\right)}
       \left[\frac{\tau_1}{v_t}\left(\gamma-1-c_s^2\right) + c_s^2\tau_3\right].
\label{eq19}
\end{eqnarray}
Note, however, that {\it all} quantities defined in eq. (\ref{eq19}) can 
finally be reduced to an algebraic expression in $r_c$ with 
real coefficients that are functions of \eker. Hence
$\left(du/dr\right)_{\rm r=r_c}$ is found to be an algebraic expression 
in $r_c$ with constant co efficients those are non linear functions 
of \eker. Once $r_c$ is known for a set of values of \eker,
the critical slope, i.e., the space gradient for $u$ at 
$r_c$ for the advective velocity can be computed as a pure 
number, which may either be a real (for transonic accretion solution to 
exist) or an imaginary (no transonic solution may be found) number. 
The critical advective velocity gradient 
for accretion solution may be computed as 
\begin{equation}
\left(\frac{du}{dr}\right)_{\rm r=r_c}
=-\frac{\beta}{2\alpha}
{\pm}
\sqrt{\beta^2-4\alpha{\zeta}}
\label{eq20}
\end{equation}
by taking the positive sign. The negative sign corresponds to 
the outflow/self-wind solution on which we would not like to concentrate 
in this work. The critical acoustic velocity gradient 
$\left(dc_s/dr\right)_{\rm r=r_c}$ can also be computed by 
substituting the value of $\left(\frac{du}{dr}\right)_{\rm r=r_c}$ 
in eq. (\ref{eq14}) and by 
evaluating other quantities in eq. (\ref{eq14}) at $r_c$. 
\subsection{On Transonicity and Multi-transonic Solutions}
\label{4.5}
\noindent
As mentioned before, the critical point can be computed by putting the 
critical point conditions in the expression on ${\cal E}$ as expressed 
in eq. (\ref{eq11}). The number of critical points obtained is either 
one (saddle type), or three (one centre type flanked by two saddle 
type) depending on the particular value of \eker  used. 
Certain \eker$_{\rm mc}{\subset}${\eker} thus provides the 
multi criticality in  accretion (as well as in outflow)
solutions, where the subscript `mc' stands for multi critical. 
It is, however, to be noted that the radial Mach number at the critical 
point is a functional of \eker and is less than unity, which is 
obvious from eq. (\ref{eq17}). 
Following our previous discussions, this is the consequence of the choice of the 
geometric configuration of the accretion flow and the equation of state to 
study such flow. As mentioned earlier, the acoustic horizon 
is defined as a time-like hypersurface defined by the equation
\begin{equation}
c_s^2-u^2=0
\label{eq21}
\end{equation}
The acoustic horizon are thus the collection of the 
`sonic' points where the radial Mach number becomes unity.
Since critical points are not topologically 
isomorphic with the 
sonic points in general, acoustic horizon does not form at the critical
point, and neither $\left(du/dr\right)_{\rm r=r_c}$ nor 
$\left(dc_s/dr\right)_{\rm r=r_c}$ can be used to evaluate the 
value of the acoustic surface gravity. Had it been the case that the 
critical points would be identical with the sonic points, the acoustic 
surface gravity could easily be evaluated by taking $r_c$ to be $r_h$, 
the radius of the acoustic horizon, and 
by directly taking the value of $c_s, u, \left(du/dr\right)_{\rm r=r_c}$
and $\left(dc_s/dr\right)_{\rm r=r_c}$ as evaluated at $r_h$. One thus understands that 
$r_h$ is actually a sonic point located on the combined integral solution of 
eq. (\ref{eq14}) and eq. (\ref{eq15}). For inviscid flow, a physically 
acceptable transonic solution can be realized to pass through a saddle type 
sonic point, resulting the hypothesis that every saddle type critical 
point is accompanied by its sonic point but no centre type critical point
has its sonic counterpart. For axisymmetric flow in the Kerr metric,
a multi-critical flow is thus a theoretical 
abstraction where three critical points are obtained as a mathematical 
solution of the energy conservation equation (through the critical point
condition), whereas a multi-transonic flow is a practically realizable 
configuration where accretion solution passes through two different saddle 
type sonic points. One should, however, note 
that a smooth accretion solution can never encounter more than one regular 
sonic points, hence no continuous transonic solution exists which passes 
through two different acoustic horizons. The only way the multi transonicity 
could be realized as a combination of two different otherwise smooth solution
passing through two different saddle type critical (and hence sonic) points and 
are connected to each other through a discontinuous shock transition. 
Such a shock has to be stationary and will be located in between two sonic
points. For certain \eker$_{\rm nss}{\subset}$\eker$_{\rm mc}$, where
`nss' stands for no shock solution, three 
critical points (two saddle embracing a centre one) 
are routinely obtained but no stationary shock 
forms. Hence no multi transonicity is observed even if the flow is 
multi-critical, and real physical accretion solution can have access 
to only one saddle type critical points (the outer one) out of the two. Thus multi
critical accretion and multi transonic accretion are not topologically 
isomorphic in general. A true multi-transonic flow can only be 
realized for
\eker$_{\rm ss}{\subset}$\eker$_{\rm mc}$, where 
`ss' stands for `shock solution', if the following criteria for forming a 
standing shock (the relativistic Rankine-Hugoniot condition,
see, e.g., \cite{abr06,das07} 
for further detail including the formulation) gets satisfied
\begin{eqnarray}
\left[\left[{\rho}u\Gamma_{u}\right]\right]=0
& & \nonumber \\
\left[\left[{\large\sf T}_{t\mu}{\eta}^{\mu}\right]\right]=
\left[\left[(p+\epsilon)v_t u\Gamma_{u} \right]\right]=0
& & \nonumber \\
\left[\left[{\large\sf T}_{\mu\nu}{\eta}^{\mu}{\eta}^{\nu}\right]\right]=
\left[\left[(p+\epsilon)u^2\Gamma_{u}^2+p \right]\right]=0
\label{eq22}
\end{eqnarray}
where $\Gamma_u=1/\sqrt{1-u^2}$ is the Lorentz factor.


The overall procedure to 
compute the value of the analogue surface gravity, is, however, 
a bit involved. For a set of \eker, $r_c,
u{\vc},{c_{s}}_{\vc},\left(du/dr\right)_{\rm 
r=r_c}, \left(dc_s/dr\right)_{\rm r=r_c}$ can be calculated. These 
values are then used as the initial values to find the 
integral solution for eq. (\ref{eq14} - \ref{eq15}). Along the integral solution, 
the radial Mach number $M$ is computed at every $r$ until one obtains the 
specific value of $r$ for which $M$ becomes exactly equal to unity. 
$r_{M=1}$ is thus equal to $r_h$. We then calculate $u,c_s,du/dr$ and
$dc_s/dr$ at $r_h$ and calculate the value of the acoustic surface 
gravity $\kappa$ using such values. The 
exact expression of $\kappa$ will be derived in the next section.
\section{Acoustic Surface Gravity in terms of \eker}
\label{sec5}
\noindent
For non-dissipative, barotropic, relativistic irrotational flow, the wave
equation describing the propagation of the acoustic perturbation can be 
described as \cite{bil99}
\begin{equation}
\frac{1}{\sqrt{-\left|G_{\mu\nu}\right|}}
\partial_\mu\left({\sqrt{-\left|G_{\mu\nu}\right|}}G_{\mu\nu}\right)
\partial_\nu{\varphi}=0
\label{eq23}
\end{equation}
by linearizing the relativistic Euler equation and the continuity equation
around some steady background. $\varphi$ in eq. (\ref{eq23}) represents the 
low amplitude linear perturbation of the velocity potential around the 
aforementioned steady background. The acoustic metric tensor 
and its inverse may be defined as
\begin{equation}
{G}_{\mu\nu} =\frac{\rho}{h c_s}
\left[g_{\mu\nu}+(1-c_s^2)v_{\mu}v_{\nu}\right];
\;\;\;\;
G^{\mu\nu} =\frac{h c_s}{\rho}
\left[g^{\mu\nu}+(1-\frac{1}{c_s^2})v^{\mu}v^{\mu}
\right] ,
\label{eq24}
\end{equation}
where $h$ is the relativistic enthalpy and $g_{\mu\nu}$ is the 
stationary background metric. 

For the steady axisymmetric fluid flow considered in this work, the 
acoustic metric is assumed to be stationary and any displacement along the 
projection of the flow velocity on ${\Sigma_v}$ is assumed to be an isometry, 
where ${\Sigma_v}$ is the hypersurface of constant $v$. The acoustic 
ergo-region in such configuration may be defined as the region where the 
stationary Killing vector becomes spacelike, i.e., any supersonic 
region is an ergo region. The stationary limit surface ${\Sigma_{c_s}}$ 
is realized as the boundary of the ergo region as defined by the equation 
\begin{equation}
v^2-c_s^2=0
\label{eq25}
\end{equation}
In terms of the acoustic metric tensor, ${\Sigma_{c_s}}$ is defined 
by the condition $G_{tt}=0$. The acoustic horizon can now be defined 
as a timelike hypersurface satisfying the criteria
\begin{equation}
\frac
{
\left(\eta^\mu v_\mu\right)^2}
{
\left(\eta^\mu{v_\mu}\right)^2 +\eta^\mu\eta_\mu} - c_s^2 =0
\label{eq26}
\end{equation}
where $\eta^\mu$ is the unit normal to the horizon. 

For axially symmetric accretion flow, the acoustic horizon does not 
co incide with the stationary limit surface in general, except where the three
velocity component perpendicular to the horizon is considered to be the 
flow velocity (as measured by a co rotating 
observer) of interest. One thus identifies the advective velocity $u$ 
on the equatorial plane with 
$
\left(\left(\eta^\mu v_\mu\right)/
\sqrt{\left(\eta^\mu{v_\mu}\right)^2 +\eta^\mu\eta_\mu}\right)$, and the 
acoustic horizon can essentially be defined by eq. (\ref{eq21}) as explained 
earlier. This further confirms that the sonic horizon (the 
collection of all the points where $M=1$) is the acoustic horizon.  

One now constructs a Killing vector $\chi^\mu = \xi^\mu+\Omega{\phi^\mu}$
where the Killing vectors $\xi^\mu$ and $\phi^\mu$ are the two 
generators of the temporal (constant ${\cal E}$) and axial (axisymmetric 
flow) isometry groups, respectively. When $\Omega$ is computed at $r_h$, 
$\chi^\mu$ becomes null on the transonic surface. The norm of the 
Killing vector $\chi_\mu$ may be 
computed as 
\begin{equation}
\sqrt{\chi^\mu{\chi_\mu}}
=\sqrt{\left(g_{tt}+2\Omega{g_{t\phi}}+\Omega^2{g_{\phi\phi}}\right)}
=\frac{\sqrt{{\Delta}B}}{g_{\phi{\phi}}+{\lambda}g_{t{\phi}}}
\label{eq27}
\end{equation}
With this, eq. (\ref{eq2}) takes the form
\begin{equation}
\kappa=\left|
\frac
{r
\sqrt
{
\left(r^2-2r+a^2\right)
\left(g_{\phi\phi}+2\lambda{g_{t\phi}}+\lambda^2{g_{tt}}\right)
}
}
{
{\sqrt{g_{rr}}}
\left(1-{c_s}^2\right)
\left({r^3+a^2r+2a^2-2\lambda{a}}\right)}
\left(\frac{du}{dr}-\frac{dc_s}{dr}\right)\right|_{\rm r=r_h}
\label{eq28}
\end{equation}
A knowledge of $u,c_s,du/dr$ and $dc_s/dr$  as evaluated at the sonic point
is thus sufficient to calculate $\kappa$ for a fixed set of values of
\eker.

\begin{figure}[h]
\includegraphics[]{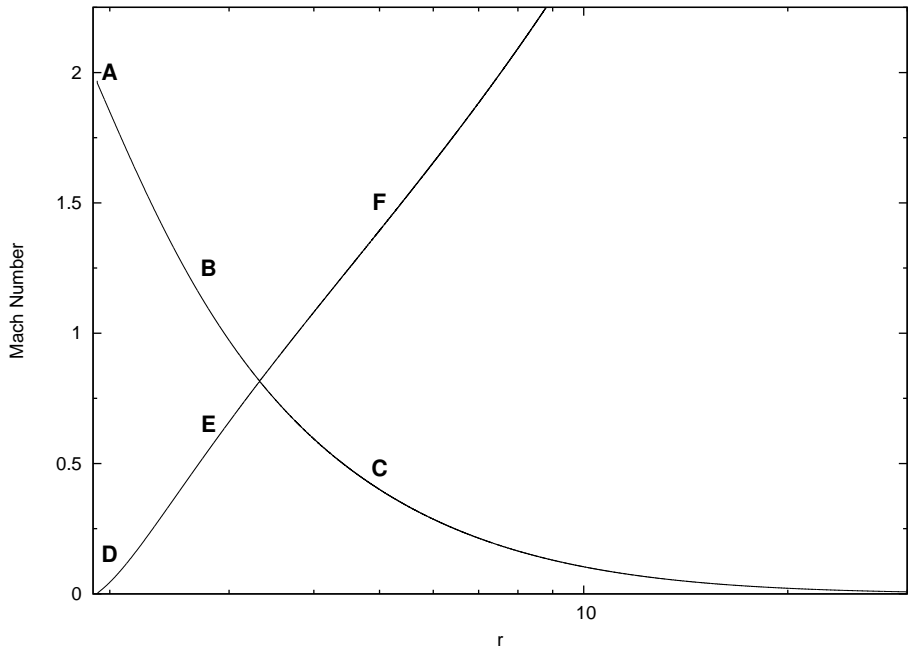}
\caption[]{Phase portrait corresponding to the mono-transonic 
accretion $ABC$ (and its associated self-wind $DEF$) characterized by 
$\left[{\cal E}=1.2,\lambda=2.0,\gamma=1.6,a=0.4\right]$. The inner type 
critical point is formed at $r=3.3276$, while the sonic horizon is
located ar $r=2.9503$, both in units of $GM_{\rm BH}/c^2$. See 
text for further details. }
\label{fig1}
\end{figure}

We now illustrate the procedure to calculate $\kappa$ for two typical 
flow configuration -- for a montransonic accretion solution passing 
through the inner type sonic point and a multi-transonic accretion 
solution with stationary shock. In figure 1, $ABC$ is the integral solution 
for transonic accretion obtained by simultaneously solving eq. (\ref{eq14} -
\ref{eq15}), and $DEF$ is the integral solution representing the `self
wind'. The radial Mach number is plotted along the $Y$ axis and the 
logarithmic radial distance measured from the black hole event 
horizon (along the equatorial plane) has been plotted along the $X$ 
axis. The phase portrait is obtained for 
$\left[{\cal E}=1.2, \lambda=2.0, \gamma=1.6, a=0.4\right]$. The 
intersection of $ABC$ and $DEF$ is essentially the critical point  $r_c$
located at $3.3276$ in units of $GM_{/rm BH}/c^2$. The critical point $r_c$
 is obtained by 
putting the critical point condition as obtained from eq. (\ref{eq17}) in 
the energy conservation condition as stated in eq. (\ref{eq11}), and
by solving it numerically for $r_c$. Such $r_c$ and its corresponding 
functions are then substituted back to eq. (\ref{eq17}) to obtain the 
critical velocity and the critical sound speed. The critical space 
gradient for $u$ and $c_s$ are then obtained from eq. (\ref{eq20}). 
$\left[u,c_s,du/dr,dc_s/dr\right]_{\rm r=r_c}$ is then used as initial 
value to obtain the integral solution of eq. (\ref{eq14} - \ref{eq15}).
Mach number is calculated at each point along the integral accretion solution $ABC$ and the value of $r$ for which $M=1$ is identified as $r_h$
($r_h<r_c$ for accretion for obvious reason). 
$\left[u,c_s,du/dr,dc_s/dr\right]_{\rm r=r_h}$ is then computed, and their 
values are substituted at eq. (\ref{eq28}) to obtain the surface 
gravity in terms of \eker. 

Similar exercises may be performed for the integral transonic solution for 
wind (branch $DEF$). However, in this work we are interested to study the 
acoustic geometry for the inward moving accretion flow only. $\kappa$ can 
also be computed for the monotransonic 
integral accretion solution passing through the outer type sonic point. 
The acoustic surface gravity is way much higher for solutions passing through the 
inner type sonic point compared to that passing through the outer 
type sonic point\footnote{For a full classification of the integral 
curves passing through various categories of sonic points, see, e.g., Das \& 
Czerny 2012.}. By varying any of the four parameters 
${\cal E},\lambda,\gamma$ and $a$, one can study the dependence of the 
acoustic surface gravity on the flow properties and on the property of 
the back ground space time itself.

\begin{figure}[h]
\includegraphics[]{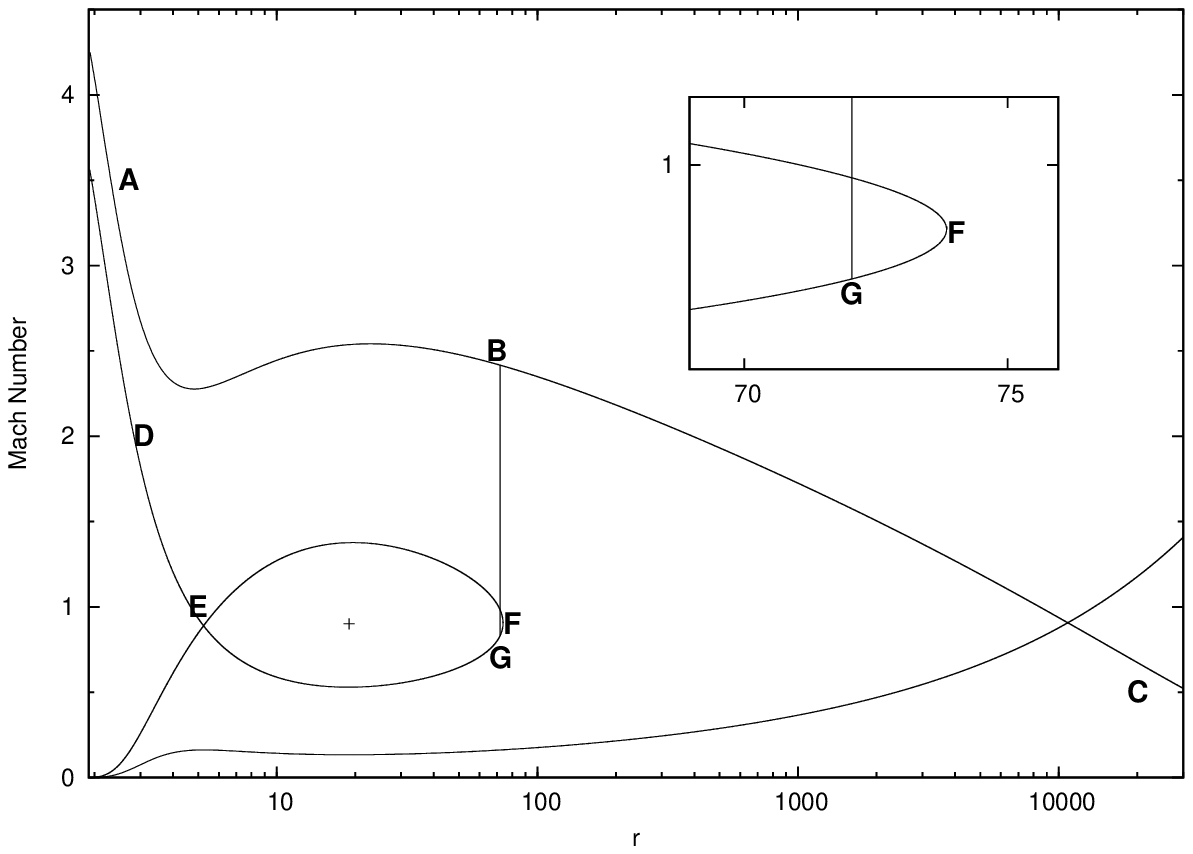}
\caption[]{Phase portrait for the multi-transonic shocked accretion 
characterized by $\left[{\cal E}=1.00001,\lambda=2.6,\gamma=1.43,a=0.4\right]$.
The vertical line $BG$ demonstrates the shock transition. The inner, middle
and the outer critical points are located at $r=5.2398$, $r=18.9471$  and $r=10829.8325$, respectively,
in units of $GM_{\rm BH}/c^2$. The inner and the outer sonic horizon
are located at $r=4.6775$ and $r=8449.4640$, respectively.
The shock is located at $r=72.01$. The middle critical point is indicated by the symbol '+'.}
\label{fig2}
\end{figure}

In figure 2, a multi-transonic flow with stationary shock has been 
depicted for $\left[{\cal E}=1.00001, \lambda=2.6, \gamma=1.43, a=0.4\right]$.
$ABC$ is the integral accretion solution passing through the outer 
critical point $r_c^{\rm out}$ and $DEGF$ is the lower half of the 
homoclinic orbit representing the transonic solution passing through the 
inner critical point $r_c^{\rm in}$. The vertical line $BG$ represents the shock 
transition. The shock location is obtained by solving the relativistic 
Rankine Hugoniot condition as stated in eq. (\ref{eq22}). $CBGED$ is the 
combined multi-transonic integral solutions with stationary shock. 
Starting from 
$\left[u,c_s,du/dr,dc_s/dr\right]_{\rm r=r_c^{out}}$ and 
$\left[u,c_s,du/dr,dc_s/dr\right]_{\rm r=r_c^{in}}$, 
one calculates 
$\left[u,c_s,du/dr,dc_s/dr\right]_{\rm r=r_h^{out}}$ and 
$\left[u,c_s,du/dr,dc_s/dr\right]_{\rm r=r_h^{in}}$, respectively. The 
corresponding surface gravity $\kappa_{out}$ and $\kappa_{in}$ 
are then computed for the 
outer and the inner acoustic horizons $r_h^{\rm out}$ and $r_h^{\rm in}$,
respectively. The ratio of the surface gravities at the inner acoustic horizon 
to that of the outer acoustic horizon is then computed as 
\begin{equation}
\kappa_{\rm io}=\frac{\kappa_{\rm in}}{\kappa_{\rm out}}
\label{eq29}
\end{equation}
At the shock, the pre and the post shock Mach numbers $M_-$ and $M_+$
are the Mach numbers calculated at the shock location for the accretion 
solutions passing through the inner and the outer sonic points,
respectively, and the shock strength $R_M=M_-/M_+$ can be 
computed for a fixed set of \eker. By varying any one of the 
four parameters ${\cal E},\lambda,\gamma$ or $a$, the dependence 
of the ratio of two acoustic surface gravities on the shock strength 
can the studied by plotting $\kappa_{io}$ as a function of $R_M$. 
Similarly, the dependence of $\kappa_{io}$ on ${\cal E},\lambda,\gamma$ 
and $a$ can be studied for the multi-transonic configuration.
\section{Dependence of $\kappa$ on \eker and on Various Shock Related Variables}
\label{sec6}
\noindent
In this section we will illustrate how the properties of the background 
space time (the properties of the black hole metric manifested apparently 
through the Kerr parameter $a$) influences the properties of the 
perturbed manifold (the properties of the acoustic metric as 
manifested through the measure of the acoustic surface gravity
$\kappa$, or the ratio $\kappa_{\rm io}$ of such $\kappa$'s at the 
inner and the outer acoustic horizon, respectively) for transonic 
accretion with and without shock. We will also investigate how the 
dynamical (manifested through the dependence of 
$\kappa$ on $\left[{\cal E},\lambda\right]$) and the thermodynamic
(manifested through the dependence of $\kappa$ on $\gamma$)properties 
of the fluid
flow influences the characteristic features of the acoustic geometry. 
We will consider both shocked and non shocked accretion of the prograde 
as well as the retrograde flow. 

As mentioned earlier, we would like to investigate the variation of $\kappa$ with each 
of the parameters $\left[{\cal E},\lambda,\gamma,a\right]$. It is to be 
understood that while studying the dependence of $\kappa$ on any one 
of the aforementioned four parameters, say ${\cal E}$, $\lambda,\gamma$ and
$a$ has to be kept fixed for the entire range of ${\cal E}$ for which 
the value of $\kappa$ has to be computed. Since for a limited non 
linear range of \eker  one obtains a multi-transonic flow with shock,
a continuous range of all parameters can not be explored for such flow.
Whereas for the mono transonic flow, a wide range of parameters, sometimes
even the entire range of the allowed values corresponding to such 
parameters, may be explored. We will illustrate this issue in the 
subsequent section in greater detail.  
\subsection{Black Hole Spin Dependence of $\kappa$}
\label{sec6.1}
\noindent
Astrophysical black holes can 
posses a wide range of spin angular momentum 
\cite{mil09,zio10,tch10,rey11,mcc11,mar11}. 
We would thus 
like to cover a sufficiently large domain of $a$, from slowly rotating 
to the near extremally rotating Kerr black holes, while studying the spin
dependence of the acoustic surface gravity for the prograde flow. 
However, multi-transonic flow characterized 
by a single set of $\left[{\cal E},\lambda,\gamma\right]$
can not cover the entire stretch of the 
aforementioned domain since the shock does not form 
(i.e., eq. (\ref{eq22}) does not get satisfied) for all values of $a$
for a single set of $\left[{\cal E},\lambda,\gamma\right]$. 
In this work we explore three different ranges
of the Kerr parameters for the multi-transonic prograde flow. Flow with 
different ranges of $a$ as are characterized by 
three different $\lambda$ (=2.6, 2.17 and 2.01 for the 
three different ranges of $a$), for the uppermost, 
middle and the lower panel, respectively) 
for fixed set of values of $\left[{\cal E}=1.00001,
\gamma=1.43\right]$.

From recent theoretical and observational findings, the 
relevance of the counter rotating accretion in black hole 
astrophysics is being increasingly evident
\cite{dau10,nix11,tch12}.
It is thus instructive to study whether the characteristic features of the acoustic 
geometry remains invariant for a direct `spin flip'. In other words, 
whether the $\kappa$ -- $a$ profile changes significantly when the initial 
boundary condition is altered from \eker  to 
$\left[{\cal E},\lambda,\gamma,-a\right]$. It is obvious that such 
investigation can not be performed for multi-transonic accretion since 
shock condition can never be satisfied for a certain set of 
\eker  as well as for $\left[{\cal E},\lambda,\gamma,-a\right]$, 
i.e., for flows characterized by exactly the same value of 
$\left[{\cal E},\lambda,\gamma,\right]$ and magnitude wise same but 
direction wise different 
values of $a$. This is 
because the region of the four dimensional parameter space
spanned by \eker  responsible for the shock formation does not allow 
any parameter degeneracy. Monotransonic accretion solutions, however, 
is not constrained by this issue, and one can obtain such solutions 
when \eker  gets directly altered to 
$\left[{\cal E},\lambda,\gamma,-a\right]$. 
However, dependence of $\kappa$ on $a$ for multi-transonic shocked
retrograde flow has also been performed for in this work, see subsequent sections 
for further details.
\subsubsection{Monotransonic Accretion}
\label{sec6.1.1}

\begin{figure}[h]
\includegraphics[]{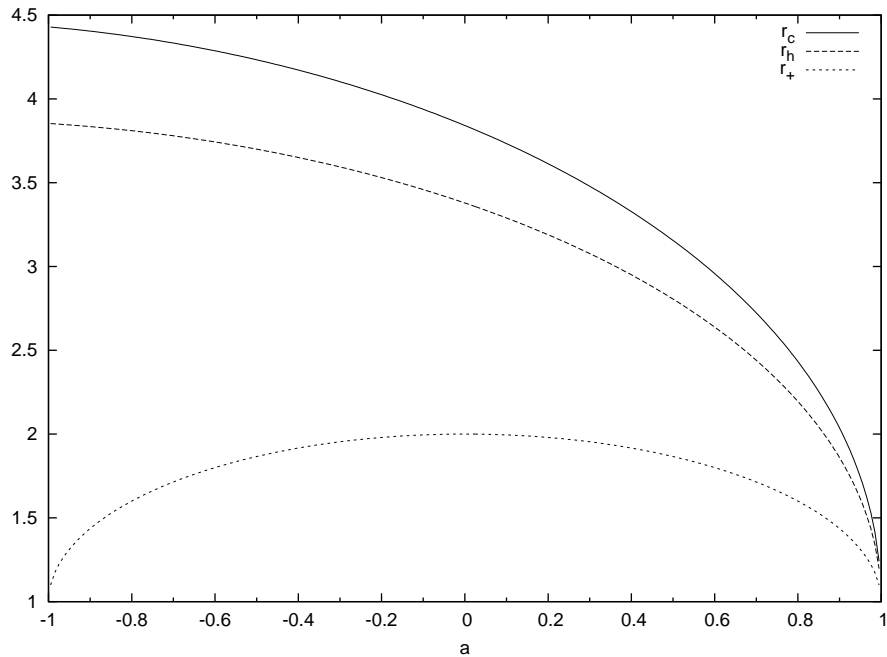}
\caption[]{The location of the acoustic horizon $r_h$ (dashed line), the 
corresponding critical point $r_c$ (solid line) and $r_+$ (dotted line), 
all in units of $GM_{\rm BH}/c^2$,
as a function of the black hole spin for mono-transonic accretion 
characterized by $\left[{\cal E}=1.2,\lambda=2.0,\gamma=1.6\right]$.}
\label{fig3}
\end{figure}

\noindent
We first calculate the acoustic surface gravity for mono-transonic 
accretion corresponding to the phase portrait as shown in figure 1. 
$\left[{\cal E}=1.2,\lambda=2.0,\gamma=1.6\right]$ has been used to calculate
$\kappa$ 
for all values of the Kerr parameter ranges from $a=-1$ to $a=1$ to explore the 
{\it entire range} of the prograde as well as the retrograde flow. In figure 3, 
the location of the acoustic horizon $r_h$ is plotted as a function 
of the $a$. Such variation is shown by the dashed curve. In the same figure the variation of $r_c$ 
with $a$ is shown by the solid curve. As a reference, location of $r_+=1+\sqrt{1-a^2}$ as a function 
of $a$ has also been plotted in the same figure using the dotted curve. The location of the acoustic 
horizon anti-correlates with the black hole spin. $r_c$ anti-correlates with 
$a$ as well. We define ${\Delta}r_{c_s}^{\left[{\cal E},\lambda,\gamma\right]}
=\left(r_c-r_h\right)$ as the difference of the location of the critical point and the 
sonic point (acoustic horizon), respectively, for a fixed set of value of 
$\left[{\cal E},\lambda,\gamma\right]$. 
${\Delta}r_{c_s}^{\left[{\cal E},\lambda,\gamma\right]}$ non linearly 
anti correlates with $a$. For very large value of the black hole spin
for co rotating accretion, the critical point almost co incides with the sonic 
point. However, the critical point and the sonic point are never 
identical (${\Delta}r_{c_s}^{\left[{\cal E},\lambda,\gamma\right]}{\ne}0$
always) for any value of $a$.

\begin{figure}[h]
\includegraphics[]{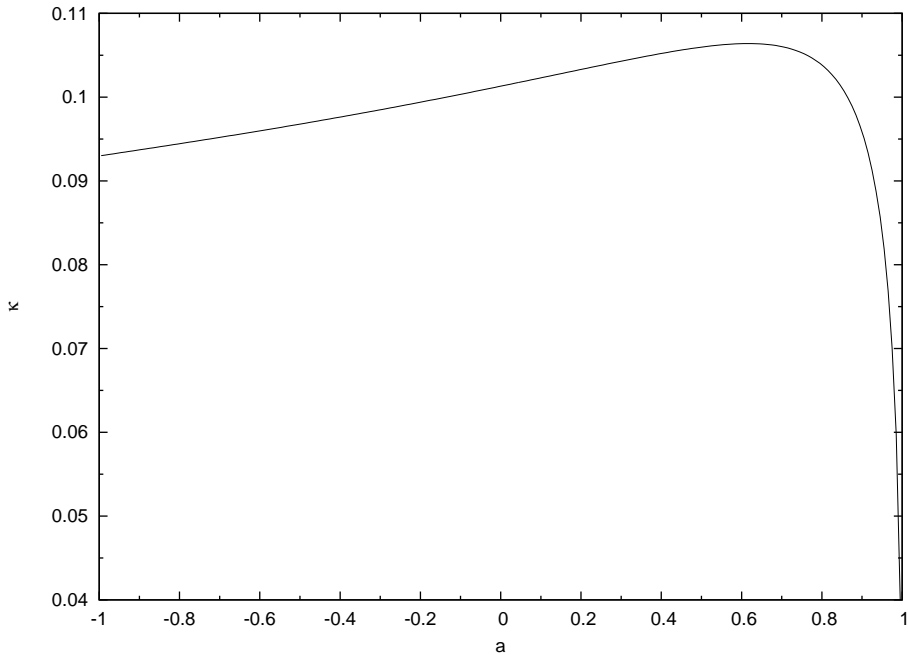}
\caption[]{The surface gravity $\kappa$ evaluated at the 
acoustic horizon as a function of black hole spin, $a$, for mono-transonic 
accretion characterized by a fixed set of 
$\left[{\cal E}=1.2, \lambda=2.0, \gamma=1.6\right]$.}
\label{fig4}
\end{figure}

In figure 4, we plot the surface gravity as a function of the black hole 
spin. For retrograde flow, $\kappa$ co relates with the black hole spin.
For prograde flow, the surface gravity initially increases non linearly 
with the black hole spin and attains a maximum value for a moderately 
high value of $a$, after which it falls off non linearly as $a$ is increased 
further. The location of the peak of the `$\kappa - a$' graph on the abscissa,
i.e., the value of $a$ for $\kappa=\kappa_{\rm max}^{\left[{\cal E},
\lambda,\gamma\right]}$ depends on the choice of 
$\left[{\cal E},\lambda,\gamma\right]$. We find that 
$a_{\rm max}^{\left[{\cal E},\lambda,\gamma\right]}$ 
nonlinear anti-correlates with $\left[{\cal E},\lambda,\gamma\right]$ whereas
$\kappa_{\rm max}^{\left[{\cal E},\lambda,\gamma\right]}$ non-linearly 
correlates with $\left[{\cal E},\lambda,\gamma\right]$.
For the mono-transonic flow $\kappa$ in general co-relates with 
$\left[{\cal E},\lambda,\gamma\right]$. This is because $r_h$ anti-correlates 
with $\left[{\cal E},\lambda,\gamma\right]$, and closer the acoustic horizon 
forms to the actual black hole event horizon, higher is the value of the 
associated surface gravity. 
\noindent
It is interesting to note that similar non-monotonic behaviour of 
`$\kappa - a$' dependence for the prograde flow is observed for 
multi-transonic shocked accretion as well. The surface gravity computed at
the inner acoustic horizon for such flow exhibits a maximum for a certain 
high value of the black hole spin. Such results are presented in the
subsequent sections.
\subsubsection{Multi-transonic Shocked Accretion}
\label{sec6.1.2}
\noindent
The typical phase diagram for a representative 
flow topology has already been shown in figure 2. For prograde flow, three 
different ranges of the black hole spin has been studied for same values of
$\left[{\cal E}=1.00001,\gamma=1.43\right]$ but for three 
different values of $\lambda$ - $\lambda=2.6$ for $a$ ranging from 
$\left[a=0.21067333~{\rm to}~a=0.47828004\right]$,
$\lambda=2.17$ for  $a$ ranging from
$\left[a=0.85614997~{\rm to}~a=0.92420954\right]$, and 
$\lambda=2.01$ for $a$ ranging from 
$\left[a=0.96821904~{\rm to}~a=0.98999715\right]$, respectively. 
For the retrograde flow, $\left[{\cal E}=1.00001,\lambda=3.3,\gamma=1.4\right]$
has been used to study the range of $a$ from 
$\left[a=-0.21~{\rm to}~a=-.661200\right]$. Note that the values 
of $\left[{\cal E},\lambda,\gamma\right]$ used here are just some 
representative set of values for which the shock forms for a substantially 
large range of $a$. Any other set of values of 
$\left[{\cal E},\lambda,\gamma\right]$ for which a reasonable range of Kerr 
parameter satisfies the relativistic Rankine Hugoniot condition can 
be used as well to study the dependence of the acoustic surface gravity on 
the black hole spin.  

\begin{figure}[h]
\includegraphics[]{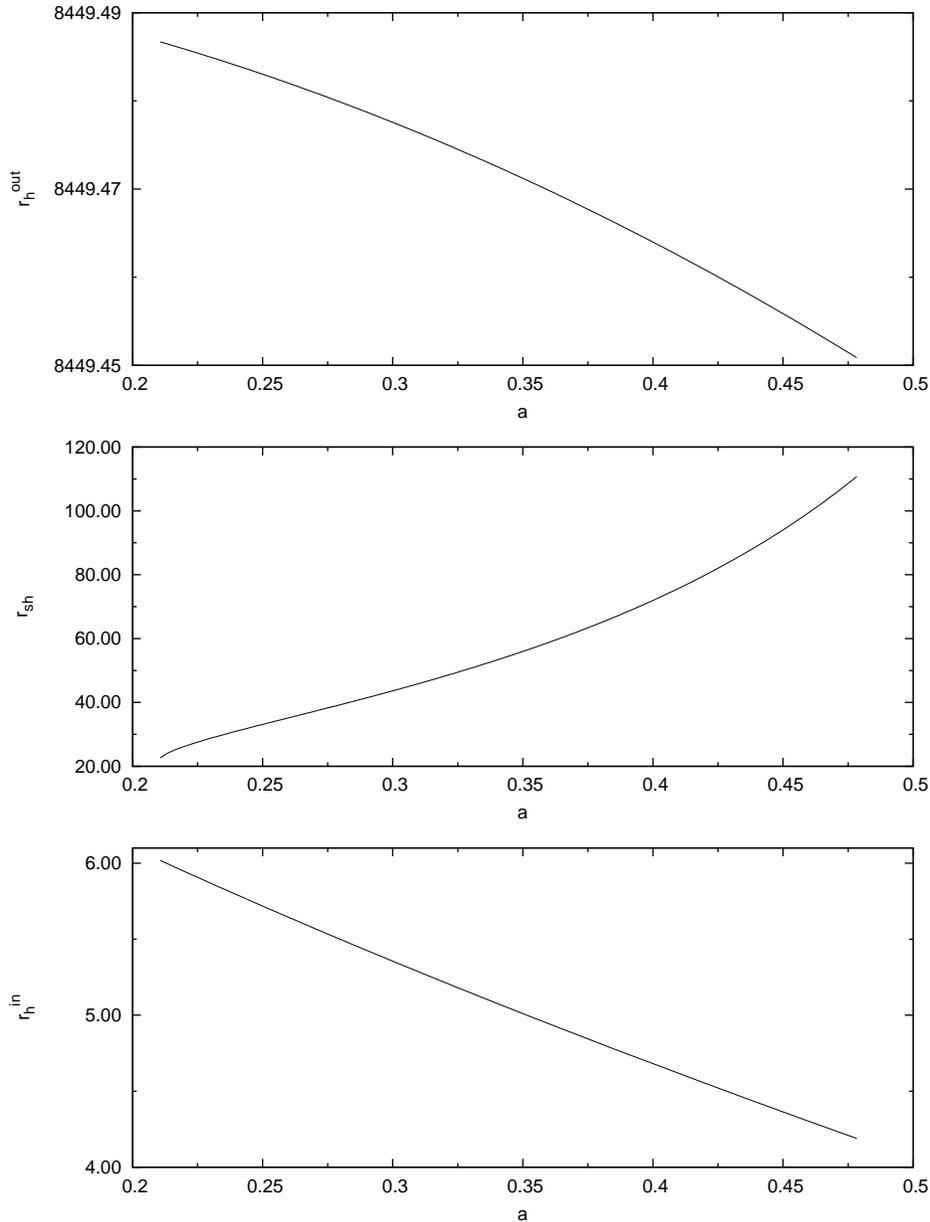}
\caption[]{The location of the 
the outer acoustic horizon $r_{h}^{\rm out}$ (upper panel), 
the stationary shock $r_{\rm sh}$ (mid panel), 
and inner acoustic horizon  $r_{h}^{\rm in}$ (lowermost panel)
as a function of black hole spin for multi-transonic shocked flow
characterized by 
$\left[{\cal E}=1.00001, \lambda=2.6, \gamma=1.43\right]$.}
\label{fig5}
\end{figure}

In figure 5, we plot the value of the location of the outer acoustic horizon
(uppermost panel), the shock location (mid panel) and the inner
acoustic horizon (lowermost panel) as a function of the Kerr parameter $a$ for 
prograde flow characterized by 
$\left[{\cal E}=1.00001,\lambda=2.6,\gamma=1.43\right]$. The location 
of the inner and the outer acoustic 
horizon non linearly anti-correlates with 
$a$, whereas the shock location non linearly co relates with $a$. 
The variation of the outer acoustic horizon is much insensitive on $a$,
which is probably 
expected since the outer horizon forms at a very large 
distance and $a$ being the inner boundary condition, does not play any 
significant role in influencing the variation of any accretion related
quantity at that length scale. On the other hand, variation of the inner acoustic horizon 
as well as of the shock location are considerably sensitive to the 
black hole spin. Similar variation of the location of the inner acoustic 
horizon $r_h^{in}$, the outer acoustic horizon $r_h^{\rm out}$ and the 
shock location $r_{sh}$ on the black hole spin can be studied for 
other set of ranges of $a$ for both prograde as well as for the 
retrograde flow. The overall 
`$\left[r_h^{\rm in},r_{sh},r_h^{\rm out}\right] - a$' profile remains more or less 
the same for any range of $a$ used -- although the numerical 
values of $\left[r_h^{in},r_{sh},r_h^{out}\right]$ differs for the different 
ranges of $a$ as characterized by by different set of 
$\left[{\cal E},\lambda,\gamma\right]$. 

An acoustic black hole horizon is formed at the inner acoustic horizon 
as well as at the outer acoustic horizon, and an acoustic white hole is formed 
at the shock location. $\kappa_{\rm io}=\kappa_{\rm in}/\kappa_{\rm out}$ can be 
calculated by obtaining the value of $\kappa_{\rm in}$ at $r_h^{\rm in}$ and 
$\kappa_{out}$ at $r_h^{out}$. However, it is observed that the 
value of $\kappa_{\rm out}$ is of the order of $10^5$ times less than 
that of $\kappa_{\rm in}$. Also the variation of $\kappa_{\rm out}$ is 
quite insensitive on $a$. This happens because the outer acoustic horizon 
forms at a very large distance from $r_+$ and the Kerr parameter is 
essentially an inner boundary condition. Hence the 
`$\kappa_{\rm io} - a$' profile is almost identical with the scaled version of 
the `$\kappa_{\rm in} - a$' profile.

\begin{figure}[h]
\includegraphics[]{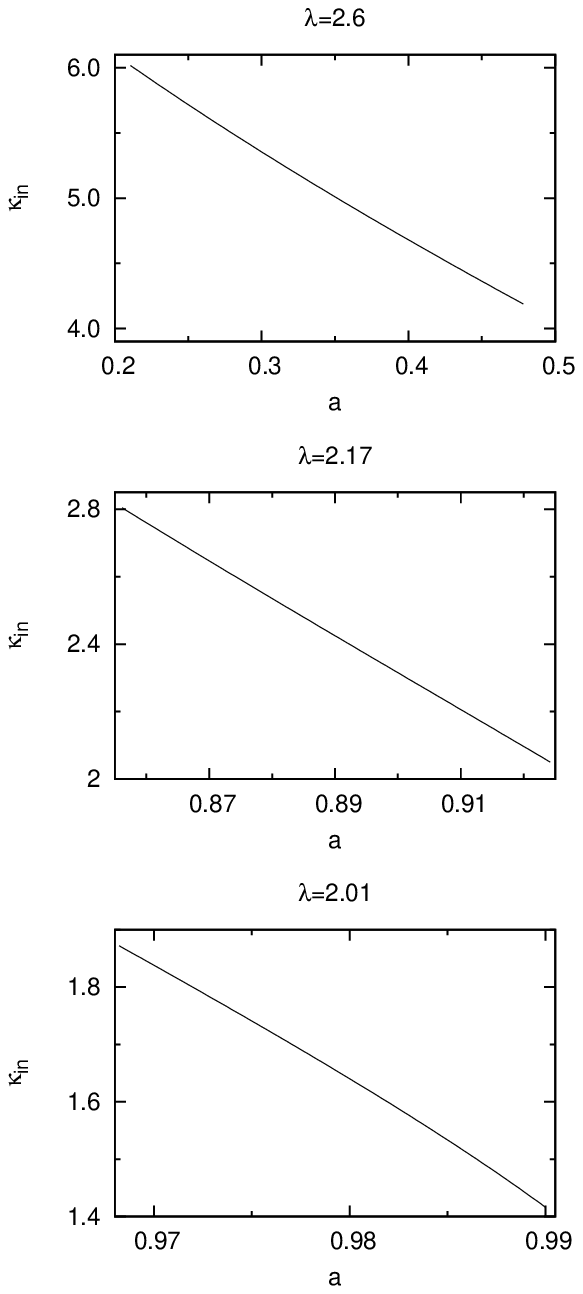}
\caption[]{The surface gravity $\kappa_{\rm in}$ evaluated at the inner 
acoustic horizon $r_h^{\rm in}$ has been plotted as a function of black hole
spin for three different ranges of the Kerr parameters. Multi-transonic 
shocked flow for a fixed set of 
values of $\left[{\cal E}=1.00001,\gamma=1.43\right]$
has been explored for three different values of the angular momentum
$\lambda=2.6$ (the uppermost panel), $\lambda=2.17$ (mid panel)
and $\lambda=2.01$ (lowermost panel) respectively, to cover the 
different ranges of the Kerr parameter. }
\label{fig6}
\end{figure}

In figure 6, we plot $\kappa_{\rm in}$ as a function of $a$ for three
different ranges of $a$ for flow characterized by same 
$\left[{\cal E},\gamma\right]$ but for three different values of $\lambda=2.6$
(uppermost panel), $\lambda=2.17$ (mid panel), and $\lambda=2.01$
(lowermost panel), respectively. As observed for the mono transonic accretion,
the surface gravity exhibits a maximum. For the multi-transonic flow,
however, the $\kappa_{\rm max}^{\left[{\cal E},\lambda,\gamma\right]}$
at the inner acoustic horizon is characterized by a very large value of 
the black hole spin.

\begin{figure}[h]
\includegraphics[]{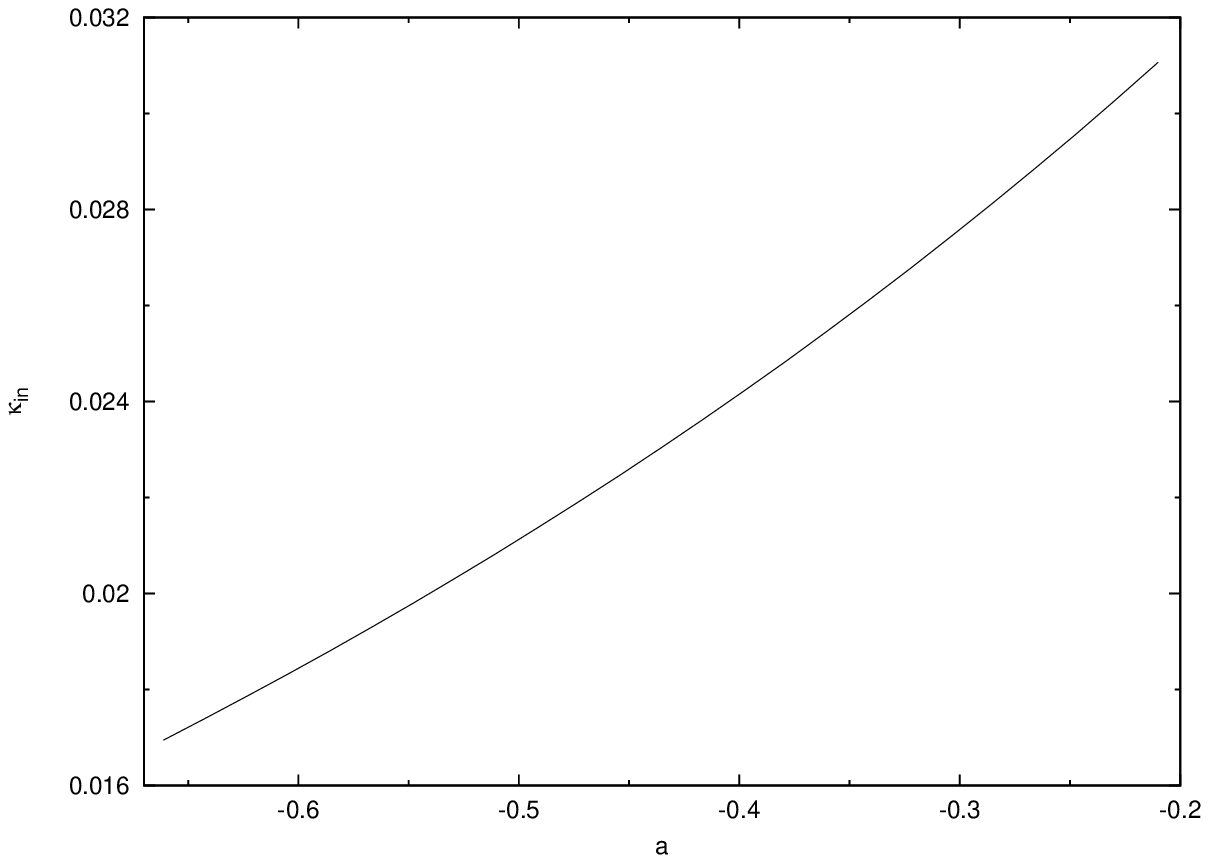}
\caption[]{The surface gravity $\kappa_{\rm in}$ evaluated at the
inner acoustic horizon for multi-transonic 
shocked retrograde accretion characterized by 
$\left[{\cal E}=1.00001,\lambda=3.3,\gamma=1.4\right]$.}
\label{fig7}
\end{figure}

In figure 7, we plot $\kappa_{\rm in}$ as a function of $a$ for 
retrograde flow characterized by 
$\left[{\cal E}=1.00001,\lambda=3.3,\gamma=1.4\right]$. 
$\kappa_{\rm in}$ non linearly anti correlates with $a$ for the retrograde flow.
\subsection{On Possible Spectral Signature of Relativistic Acoustic Geometry}
\label{sec6.2}

\begin{figure}[h]
\includegraphics[]{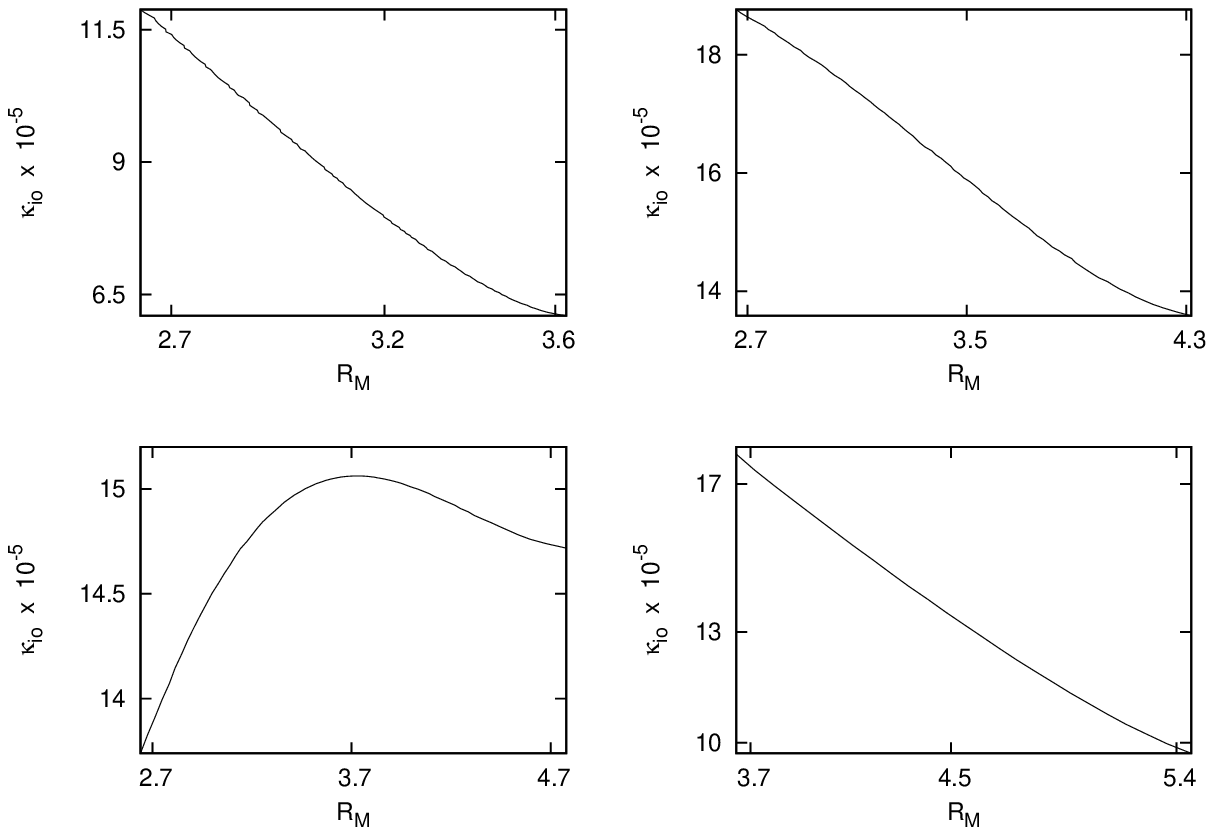}
\caption[]{$\kappa_{\rm io}$, the ratio of the surface gravities
as evaluated at the inner acoustic horizon to that as evaluated at the 
outer acoustic horizon has been plotted (scaled by a factor of $10^{-5}$)
as a function of the shock strength $R_M$ for three different ranges of the 
black hole spin for the prograde flow as characterized by 
a fixed set of $\left[{\cal E}=1.00001, \gamma=1.43\right]$ and 
three different values of $\lambda=2.6$ (top left panel),
$\lambda=2.17$ (top right panel), $\lambda=2.01$ (bottom left panel),
and for retrograde flow as characterized by 
$\left[{\cal E}=1.00001, \lambda=3.3, \gamma=1.4\right]$
(bottom right panel).}
\label{fig8}
\end{figure}

\noindent
In our attempt to make a close connection between the salient features of
the acoustic geometry and any astrophysically relevant observables, we study 
the relation between the analogue surface gravity with shock related 
accretion variables since such variables play a crucial role in determining the
spectral signature of the astrophysical black hole candidates. 
Shock formation phenomena is a discontinuous 
event somewhat equivalent to the first order phase transition. Accretion 
variables will change discontinuously at the shock location. Such accretion 
variables like density, velocity and the flow temperature provides the 
characteristic profiles of the observed spectra \cite{kat98}. 
A shocked flow will provide distinctively different 
spectra in comparison to its shock free counterpart. Corresponding 
spectra for shocked flow exhibits additional rich features 
due to the presence of the shock. The ratio of the pre and the 
post shock flow variables will determine such complex features. The flux 
distribution for such spectra for axially symmetric accretion in the Kerr
metric can be calculated in terms of such ratios (Das \& Huang, in 
preparation). In subsequent sections, we will study the dependence of the 
values of the acoustic surface gravity on pre(post) to post(pre) shock 
ratio of various relevant quantities responsible to characterize the 
spectral feature -- namely, the shock strength $R_M=M_-/M_+$, which is the 
pre to post shock Mach number ratio, the ratio of the post to 
pre shock density $R_{\rho}=\rho_+/\rho_-$ (the compression ratio) and the
temperature $R_T=T_+/T_-$, respectively, for both the prograde and the 
retrograde flow. In figure 8, we plot the ratio of the acoustic 
surface gravity $\kappa_{\rm io}=\kappa_{\rm in}/\kappa_{\rm out}$ 
(scaled by $10^{-5}$) 
with the shock strength $R_{\rm M}=M_-/M_+$ for three 
different ranges of the Kerr parameter for the prograde flow (top 
left and right and the bottom left panel) and a particular range of 
the Kerr parameter for the retrograde flow (bottom right panel). The range of
the Kerr parameters and the values of $\left[{\cal E},\lambda,\gamma\right]$
used are the same as has been used to produce the figure 6 and the figure 7.
For low to moderately high values of $a$ for the prograde flow,
$\kappa_{\rm io}$ anti-correlates with $R_M$ in general. However, for the 
substantially large value of $a$, `$\kappa_{\rm io} - a$' profile is different. 
For such range of $a$, $\kappa_{\rm io}$ initially correlates with $R_M$ 
and attains a maximum, then starts falling for larger $R_M$
non linearly as $a$ is increased further. For the retrograde flow, 
$\kappa_{\rm io}$ usually anti-correlates with $R_M$.

\begin{figure}[h]
\includegraphics[]{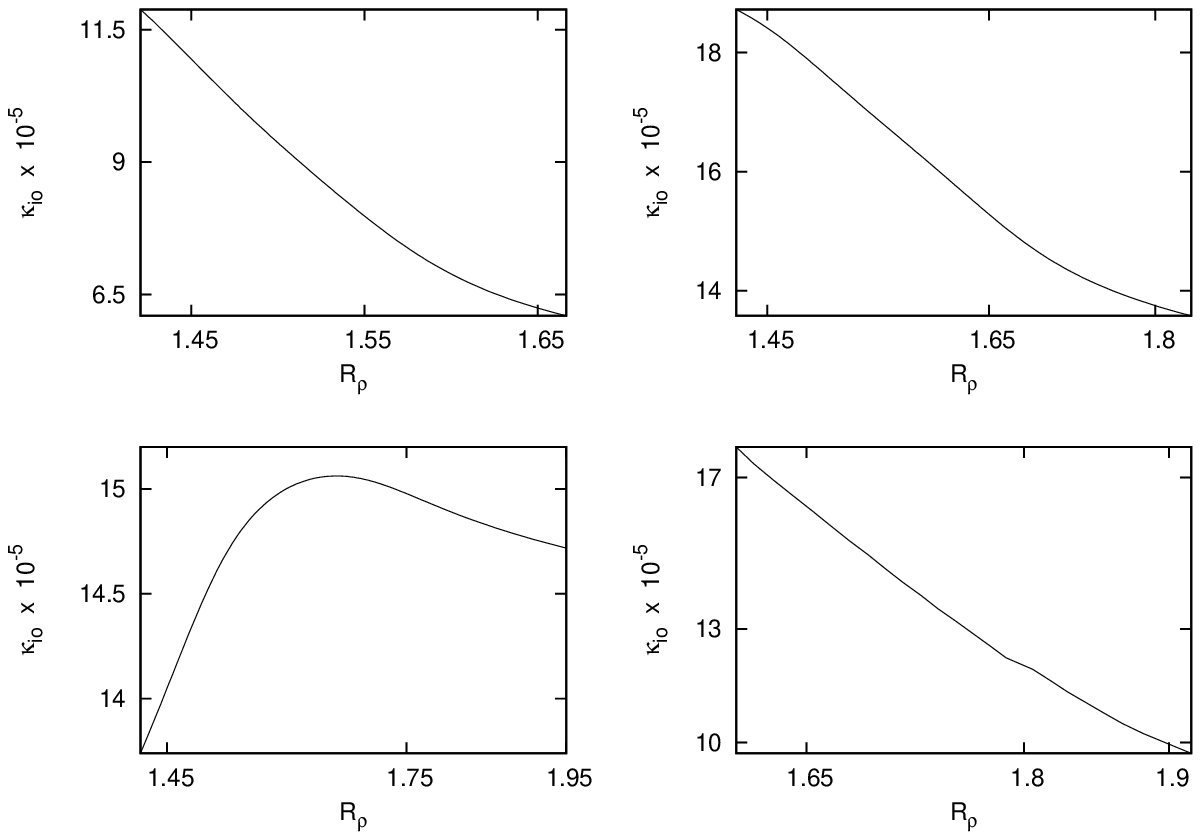}
\caption[]{$\kappa_{\rm io}$, the ratio of the surface gravities
as evaluated at the inner acoustic horizon to that as evaluated at the
outer acoustic horizon has been plotted (scaled by a factor of $10^{-5}$)
as a function of the shock strength $R_{\rho}$ for three different ranges of the
black hole spin for the prograde flow as characterized by
a fixed set of $\left[{cal E}=1.00001,\gamma=1.43\right]$ and
three different values of $\lambda=2.6$ (top left panel),
$\lambda=2.17$ (top right panel), $\lambda=2.01$ (bottom left panel),
and for retrograde flow as characterized by
$\left[{\cal E}=1.00001,\lambda=3.3,\gamma=1.4\right]$
(bottom right panel).}
\label{fig9}
\end{figure}

\begin{figure}[h]
\includegraphics[]{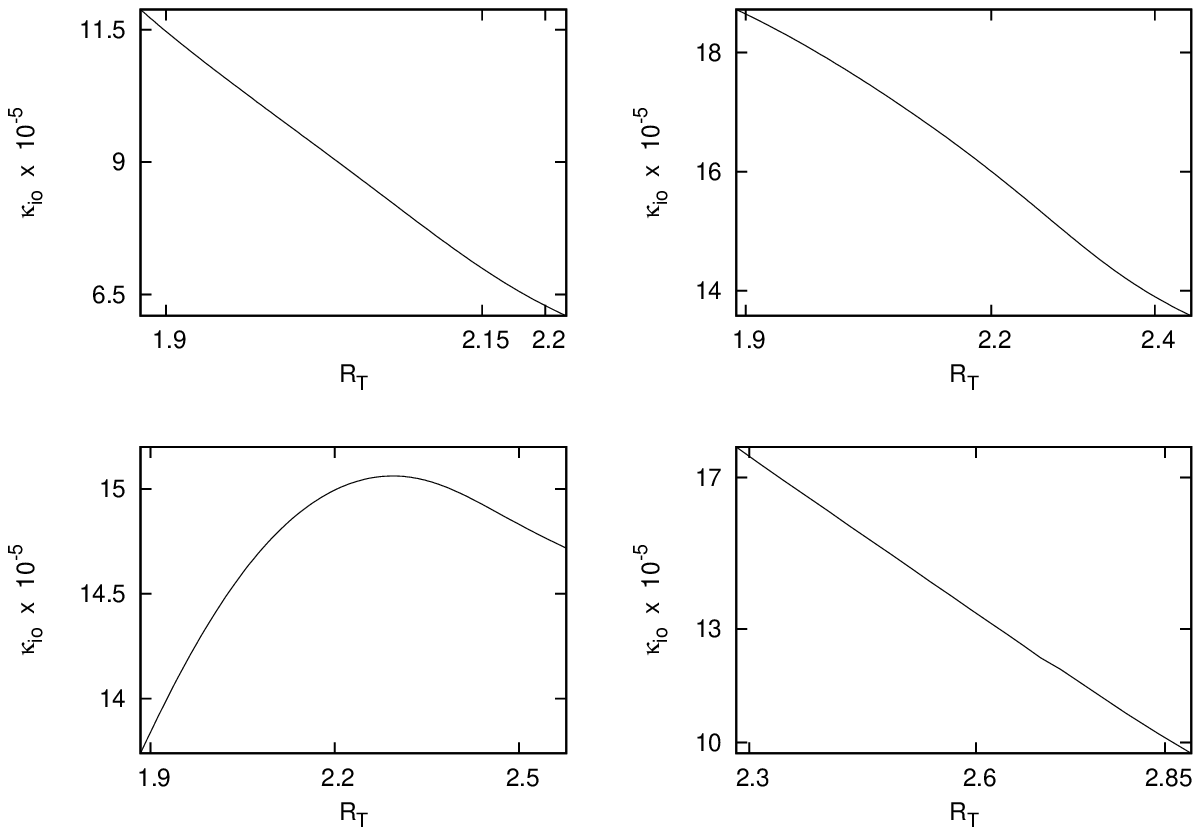}
\caption[]{$\kappa_{io}$, the ratio of the surface gravities
as evaluated at the inner acoustic horizon to that as evaluated at the
outer acoustic horizon has been plotted (scaled by a factor of $10^{-5}$)
as a function of the shock strength $R_T$ for three different ranges of the
black hole spin for the prograde flow as characterized by
a fixed set of $\left[{cal E}=1.00001,\gamma=1.43\right]$ and
three different values of $\lambda=2.6$ (top left panel),
$\lambda=2.17$ (top right panel), $\lambda=2.01$ (bottom left panel),
and for retrograde flow as characterized by
$\left[{\cal E}=1.00001,\lambda=3.3,\gamma=1.4\right]$
(bottom right panel).}
\label{fig10}
\end{figure}

Multi-transonic 
flow with shocks formed closer to the black hole are associated with 
higher values of the shock strength and larger compression ratio, and hence 
the higher value of $R_{\rm T}$ as well. Hence the `$\kappa_{\rm io} - R_{\rm M}$' 
profile should be similar with the `$\kappa_{io} - R_{\rho}$' 
and  `$\kappa_{\rm io} - R_{\rm T}$' profile. That this intuitively obvious 
conclusions follows in reality may further be formally demonstrated through 
figure 9 and figure 10 where we plot $\kappa_{\rm io}$ as a 
function of $R_{\rho}$
(figure 9) and 
$R_{\rm T}$ (figure 10) as well for the same set of values 
of the initial boundary conditions as has been used to produce figure 8. The overall 
conclusion is that except for certain span of $a$ corresponding to the near 
extremally rotating black holes, the ratio of the acoustic surface gravity 
at the inner and the outer acoustic horizon has relatively large value 
for weaker shocks formed in the prograde accretion. For retrograde accretion, 
however, the ratio $\kappa_{\rm io}$ assumes higher value for weak shocks in
general.

Since the black hole spin anti-correlates with $r_{sh}$ and correlates with
$r_h^{in}$, shocks formed at the larger distance leads to the formation
of the inner acoustic horizon relatively closer to the black hole where the
value of the acoustic surface gravity becomes substantially large.
On the other hand, shocks formed closer to the black hole are rather
strong shocks and thus produce greater compression ratio as well
as the larger value of $R_T$. Hence weaker shocks correspond to the inner
acoustic horizons located relatively closer to the black hole where
the acoustic
surface gravity assumes a higher value. On the other hand, as explained
earlier, change of the location of the outer acoustic horizon is less sensitive
to the influence of the black hole spin, hence
`$\kappa_{io} - \left[R_M,R_{\rho},R_T\right]$' variation is
effectively equivalent with the
scaled down version of `$\kappa_i - \left[R_M,R_{\rho},R_T\right]$'.
This explain why $\kappa_{io}$ anti-correlates with $R_M,R_\rho$ and $R_T$
in general. However, it is not quite clear at this stage why such
`$\kappa_{io} -- \left[R_M,R_{\rho},R_T\right]$'
profile exhibits opposite nature for some range of the black hole spin for
prograde accretion onto nearly extremely rotating holes. The inherent
complexity
of the dependence of the value of $\kappa$ on various quantities as
evaluated on the
horizon prohibits to make any conclusive remark in this case since the
analytical expression for the quantities involved are not available within the
framework of the formalism presented here.
\subsection{Dependence of $\kappa$ on $\left[{\cal E},\lambda,\gamma\right]$}
\label{sec6.3}

\begin{figure}[h]
\includegraphics[]{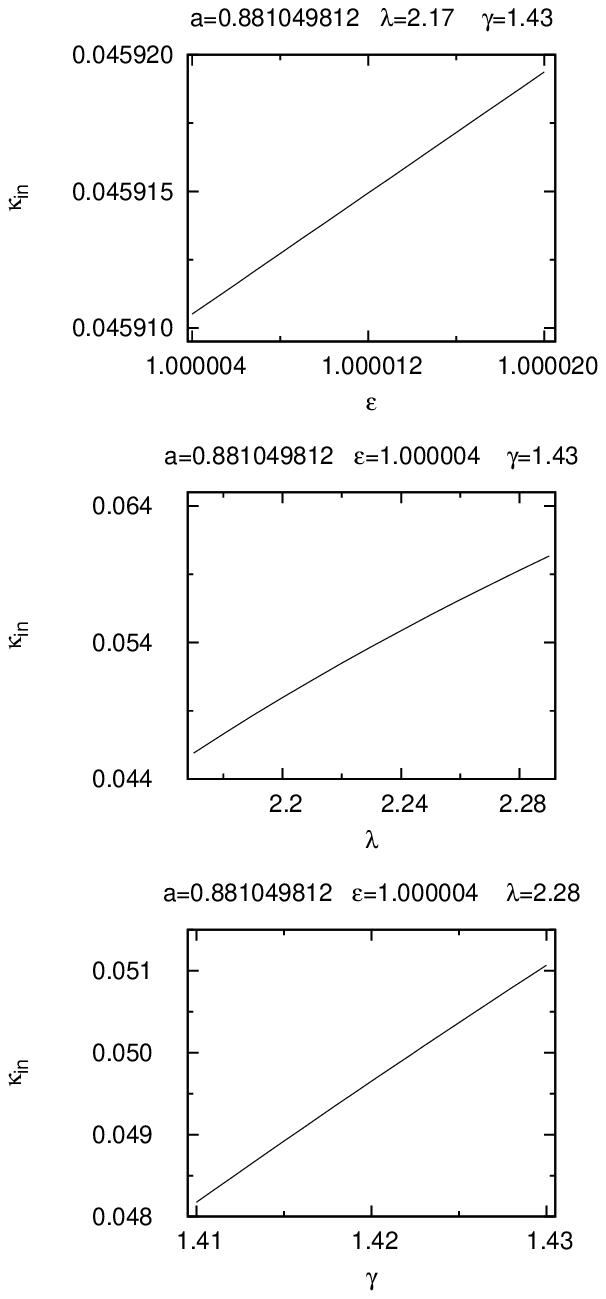}
\caption[]{For multi-transonic shocked accretion flow,
the surface gravity $\kappa_{\rm in}$ evaluated at the inner acoustic horizon 
as a function of the specific energy of the flow ${\cal E}$
(uppermost panel) for flow characterized by 
$\left[\lambda=2.17,\gamma=1.43,a=0.881049812\right]$, the 
specific angular momentum $\lambda$ (mid panel) for flow 
characterized by 
$\left[{\cal E}=1.000004,\gamma=1.43,a=0.881049812\right]$,
and of the adiabatic index of the flow 
$\gamma$ (lowermost panel) for flow characterized by 
$\left[{\cal E}=1.000004,\lambda=2.28,a=0.881049812\right]$.}
\label{fig11}
\end{figure}

\noindent
Both the inner and the outer acoustic horizon anti-correlates with 
$\left[{\cal E},\lambda,\gamma\right]$. The variation of $r_h^{in}$ 
is quite insensitive to ${\cal E}$, moderately sensitive to 
$\gamma$ and highly sensitive to $\lambda$. On the other 
hand, the variation of $r_h^{\rm out}$ is hardly sensitive to $\lambda$, 
whereas it is moderately sensitive to $\gamma$ and very much 
sensitive to ${\cal E}$. The shock location $r_{\rm sh}$ 
correlates with $\lambda$ but anti-correlates with 
$\left[{\cal E},\gamma\right]$. As usual, stronger shocks forms 
closer to the black holes. 

In figure 11, the surface gravity evaluated at the inner acoustic 
horizon for multi-transonic shocked accretion is plotted as a function of 
${\cal E}$ (uppermost panel for flow characterized by 
$\left[\lambda=2.17,\gamma=1.43,a=0.881049812\right]$), 
$\lambda$ (mid panel, for flow characterized by 
$\left[{\cal E}=1.000004,\gamma=1.43,a=0.881049812\right]$) 
and $\gamma$ (lowermost panel, for flow characterized by 
$\left[{\cal E}=1.000004,\lambda=2.28,a=0.881049812\right]$). 
$\kappa_{in}$ anti-correlates with $\left[{\cal E},\lambda,\gamma\right]$.
The variation of $\kappa_{in}$ is insensitive to ${\cal E}$,
moderately sensitive to $\gamma$ and considerably sensitive to $\lambda$.
$\kappa_{out}$ also anti-correlates with 
$\left[{\cal E},\lambda,\gamma\right]$ in general (not shown in the figure).
$\kappa_{out}$, however, is around 10$^{5}$ to 10$^6$ smaller in 
magnitude to $\kappa_{in}$. The variation of $\kappa_{out}$, however,
is insensitive to $\lambda$, moderately sensitive to $\gamma$ and
quite sensitive to ${\cal E}$. In the context to the acoustic geometry,
$\lambda$ thus somewhat controls the properties of the flow as well as 
that of the acoustic geometry close to the actual black hole, whereas 
${\cal E}$ influences the related quantities mainly at the larger 
distance. 	
\section{Discussion}
\label{sec7}
\noindent
In this work we make attempt to understand the influence of the
general relativistic background black hole
spacetime metric on the embedded perturbative manifold, i.e., on the
relativistic acoustic geometry.
To accomplish that task, we consider a specific example of classical
analogue model --
the equatorial slice of the general relativistic hydrodynamic inviscid
axisymmetric accretion onto
a spinning astrophysical black hole. We found that such accretion
configuration gives rise to  a very
interesting example of acoustic geometry where more than one acoustic
horizons may form.
We then calculate the acoustic surface gravity for such horizon(s) and
studied how the
value of such surface gravity, which is the characteristic feature of
the perturbative
manifold, depends on the spin angular momentum of the astrophysical
black holes, the Kerr parameter, which is the main characteristic
feature of the
background space-time -- as well as on various accretion parameters
which are the
characteristic features of the background fluid continuum.
Visser and Weinfurtner \cite{vis05} have demonstrated that the equatorial slice of the
Kerr geometry is equivalent to certain analogue model based on a
vortex geometry. Our
work, in some sense, is complementary to such finding. We, however,
for the first
time in the literature provided a semi analytical formalism where the
influence of
background {\it black hole space-time} has explicitly been demonstrated
on the embedded
relativistic acoustic geometry. Since almost all astrophysical black holes
are supposed to posses some degree of intrinsic rotation
\cite{mil09,dau10,zio10,tch10,nix11,rey11,mcc11,mar11,tch12}, 
the effect of the
Kerr parameter on the classical analogue model is extremely important to understand.
Our work precisely accomplished that task.

One of the most interesting aspects of acoustic surface gravity is the
existence of the
associated analogue Hawking radiation.  Acoustic horizon emits Hawking
type radiation
of thermal phonons. The corresponding analogue Hawking temperature $T_{AH}=
\kappa/{2{\pi}}$ is the characteristic temperature of such radiation measured
by an observer at infinity. Acoustic horizons explored in our work,
and its associated analogue
surface gravity are thus characterized by certain $T_{AH}$.
In this work, we have demonstrated how the measure of the acoustic surface
gravity may be associated with certain observables through the spectral feature
of the astrophysical black holes. This work, thus, makes the first
ever attempt 
to obtain the observational signature of the analogue hawking
temperature
profile for a large scale relativistic classical fluid. One, however,
should note that
$T_{AH}$ for an astrophysical black hole being too low, might be masked by
the background thermal noise, hence the accreting primordial black holes
might serve as better candidates for the purpose of the possible
measurement of the
analogue Hawking temperature.

The surface gravity can not be computed at the shock location since
$u$ and $c_s$
changes discontinuously at the shock. As a result,
$\left(du/dr\right)_{r_{sh}}$
and $\left(dc_s/dr\right)_{r_{sh}}$ diverges. The surface gravity and
the associated
analogue Hawking temperature thus becomes formally infinite at the shock.
Had it been the case that viscosity and other dissipative effects
would be included
in the system, such discontinuities in   $\left(du/dr\right)_{r_{sh}}$ and
$\left(dc_s/dr\right)_{r_{sh}}$ would have been smeared out. Under that
circumstances, $\kappa$ as well as $T_{AH}$ would have a finite
but extremely large value at the shock. This may be considered 
as a realistic manifestation of the general results 
obtained by \cite{lib00}.

In our work, however, we consider only the inviscid flow. The effect
of the viscous transport
of the angular momentum, however, is quite a subtle issue in considering the
analogue effects in black hole accretion. Thirty nine years after the
discovery of the
standard accretion disc theory \cite{nov73,sha73},
realistic modelling of viscous transonic accretion flow with
dissipation is still quite an
arduous task. From the analogue point of view, viscosity is likely to
destroy the Lorenz invariance,
and the assumptions behind constructing an acoustic geometry may not
be quite consistent for such case. Nevertheless, very large radial velocity
close to the black hole implies that the infall time scale
is substantially small compared to the viscous time scale. Large radial
velocity even at larger distances are due to the fact that the
rotational energy of the fluid is relatively low 
\cite{bel91,igu97,pro03}.
Our assumption of inviscid flow is not unjustifed from the astrophysical point
of view. One of the significant effects of inclusion of the viscosity
would be the
reduction of the angular momentum. As we demonstrate in this work, the
location of the acoustic horizon anti-correlates with $\lambda$.
Weakly rotating
flow makes the dynamical velocity gradient steeper leading to the
conclusion that
for viscous flow the acoustic horizons will be pushed further out from the
black hole and the flow would become supersonic at a larger distance for the
same set of other initial boundary conditions. The value of the surface
gravity (and
the associated analogue Hawking temperature)  anti-correlates with the
location of
the acoustic horizon. A viscous transonic accretion disc is thus expected to
produce lower value of $\kappa$ and $T_{AH}$ compared to its
inviscid counterpart.

An axially symmetric accretion flow in vertical equilibrium has been
studied in our work
where the disc height is a function of the radial distance on the
equatorial plane.
Axisymmetric accretion may also be studied for two other different
flow configurations,
namely, for disc with constant thickness and flow in conical
equilibrium where the local flow thickness
to the radial distance on the equatorial plane remains constant. For
these two flow configurations,
the critical points are isomorphic to the sonic points
\cite{abr06,nag12}. Hence the location of the acoustic
horizon and the
values of the dynamical velocity along with the sound speed and their space
derivative as well can be computed analytically without taking recourse
to the integral flow solution, hence by avoiding the process of numerical
integration. The non monotonic behaviour of the
`$\kappa - a$' profile, especially the appearance of
$\kappa_{\rm max}^{\left[{\cal E},\lambda,\gamma\right]}$ and the
dependence of of the corresponding
$a_{\rm max}^{\left[{\cal E},\lambda,\gamma\right]}$ on
$\left[{\cal E},\lambda,\gamma\right]$ will then be better understood
since we can then
directly differentiate the expression of $\kappa$ with respect to a $a$
to find out for what value of $a$ (for a fixed set of $\left[{\cal
E},\lambda,\gamma\right]$)
the $\kappa - a$ profile attains its maximum, and how such value of
$a$ changes
with the variation of $\left[{\cal E},\lambda,\gamma\right]$. We plan
to explore such analytical dependence
in our future work for axisymmetric flow in conical equilibrium.

Finally, we would like to emphasize that in the present work we did
not aim to provide
a formalism using which the phonon field generated in the system
concerned could be
quantized. To accomplish that task, one has to demonstrate that the
effective action
for the acoustic perturbation is equivalent to a field theoretical
action in curved space, and the
associated commutation and the dispersion relations should directly follow
\cite{unr95} \& \cite{sch05}. Such considerations are rather
involved and are beyond the scope of this paper. Our main motivation was rather to
apply the analogy to describe the classical perturbation of the fluid
flow in terms of
a field satisfying the wave equation in an effective geometry and to
study the relevant
consequences.

\acknowledgements
HP and IM would like to acknowledge the kind hospitality provided 
by HRI, Allahabad, India, under a visiting students research programme. TKD
would like to acknowledge the professional support provided by S. N. Bose National 
Centre for Basic Science (by offering a long term sabbatical 
visiting professor position) where part of the work has been done. 
The research of HP and HC is partially supported by the 
Natioanl Science Council of the Republic of China under the grant NSC 
99-2112-M-007-017-MY3. The research of TKD is partially supported by the
astrophysics project under the XI th plan at HRI.

\newpage


\end{document}